# A quantum circuit rule for interference effects in single-molecule electrical junctions

David Zsolt Manrique,[†,] Cancan Huang,[‡,] Masoud Baghernejad,[‡,] Xiaotao Zhao,[§,] Oday A. Al-Owaedi, ,[†,] Hatef Sadeghi,[†]

Veerabhadrarao Kaliginedi,[‡] Wenjing Hong,[‡,*] Murat Gulcur,[§] Thomas Wandlowski,[‡] Martin R. Bryce,[§,*] Colin J. Lambert,[†,*]

A quantum circuit rule for combining quantum interference (QI) effects in the conductive properties of oligo(phenyleneethynylene) (OPE)-type molecules possessing three aromatic rings was investigated both experimentally and theoretically. Molecules were of the type X-Y-X, where X represents pyridyl anchors with para (p), meta (m) or ortho (o) connectivities and Y represents a phenyl ring with p and m connectivities. The conductances $G_{XmX}$ ($G_{XpX}$) of molecules of the form X-m-X (X-p-X), with meta (para) connections in the central ring were predominantly lower (higher), irrespective of the meta, para, or ortho nature of the anchor groups X, demonstrating that conductance is dominated by the nature of QI in the central ring Y. The single-molecule conductances were found to satisfy the quantum circuit rule $G_{ppp}/G_{pmp} = G_{mpm}/G_{mmm}$. This demonstrates that the contribution to the conductance from the central ring is independent of the para versus meta nature of the anchor groups.

[†] Department of Physics, Lancaster University, Lancaster LA1 4YB, United Kingdom

[‡] Department of Chemistry and Biochemistry, University of Bern, Freiestrasse 3, CH-3012 Bern, Switzerland

[§] Department of Chemistry, Durham University, Durham DH1 3LE, United Kingdom

Studies of the electrical conductance of single molecules attached to metallic electrodes not only probe the fundamentals of quantum transport, but also provide the knowledge needed to develop future molecular-scale devices and functioning circuits.[1-9] Due to their small size (on the scale of angstroms) and the large energy gaps (on the scale of eV) transport through single molecules can remain phase coherent even at room temperature and constructive or destructive quantum interference (QI) can be utilized to manipulate their room-temperature electrical[10-13] and thermoelectrical[14,15] properties. In previous studies, it was reported theoretically and experimentally that the conductance of a phenyl ring with meta (m) connectivity is lower than the isomer with para (p) connectivity by several orders of magnitude.[16-25] This arises because partial de Broglie waves traversing different paths through the ring are perfectly out of phase leading to destructive QI in the case of meta coupling, while for para or ortho coupling they are perfectly in phase and exhibit constructive QI. (See eg equ. 8 of ref [26].) It is therefore natural to investigate how QI in molecules with multiple aromatic rings can be utilized in the design of more complicated networks of interference-controlled molecular units.

The basic unit for studying QI in single molecules is the phenyl ring, with thiol,[17,21] methyl thioether,[27] amine,[17] or cyanide[19] anchors directly connecting the aromatic ring to gold electrodes. Recently, Arroyo et al.[28,29] studied the effect of QI in a central phenyl ring by varying the coupling to various anchor groups, including two variants of thienyl anchors. However, the relative importance of QI in central rings compared with QI in anchor groups has not been studied systematically, because the thienyl anchors of Arroyo et al were 5-membered rings, which exhibit only constructive interference. To study the relative effect of QI in anchors, we examined molecules with terminal groups formed from six-membered pyridyl rings for which constructive and destructive interference is possible. We have previously reported that pyridyl rings are excellent anchor groups for attaching single-molecules to metallic electrodes, because of their well-defined binding geometry,[30] in which the nitrogen plays the role of the anchoring site. The ability to substitute pyridine either para (p), meta (m) or ortho (o) to the nitrogen offers the possibility of systematically investigating the relative importance of QI in this anchor group in molecules of the type X-Y-X, where X is a pyridyl ring and Y is a central phenyl ring. This question is rather subtle, because charge transport from the electrode into the pyridyl rings takes place via the N atom and via metal-π coupling[31] and the interplay between these two mechanisms will determine the importance and robustness of QI in the terminal[32] rings. In the present paper, our aim is to compare the effects of QI in both the terminal rings X and central ring Y of molecules of the type X-Y-X and to study the relationship between their conductances. Our results were found to satisfy the quantum circuit rule $G_{ppp}/G_{pmp} = G_{mpm}/G_{mmm}$ which demonstrates that the contribution to the conductance from the central ring is independent of the para versus meta nature of the anchor groups. Combinatorial rules for QI in molecules with other geometries, including fused rings or rings connected in parallel are discussed in refs.[33-36]

We studied the pyridyl-terminated OPE derivatives shown in Figure 1, which possess a variety of connectivities of the central ring and locations of the nitrogen in the anchor units. In group 1 the central unit Y is a phenyl ring with para connectivity, whereas the central ring of group 2 has meta connectivity. The anchor units X are pyridyl rings with their nitrogens located in either meta, ortho or para positions.

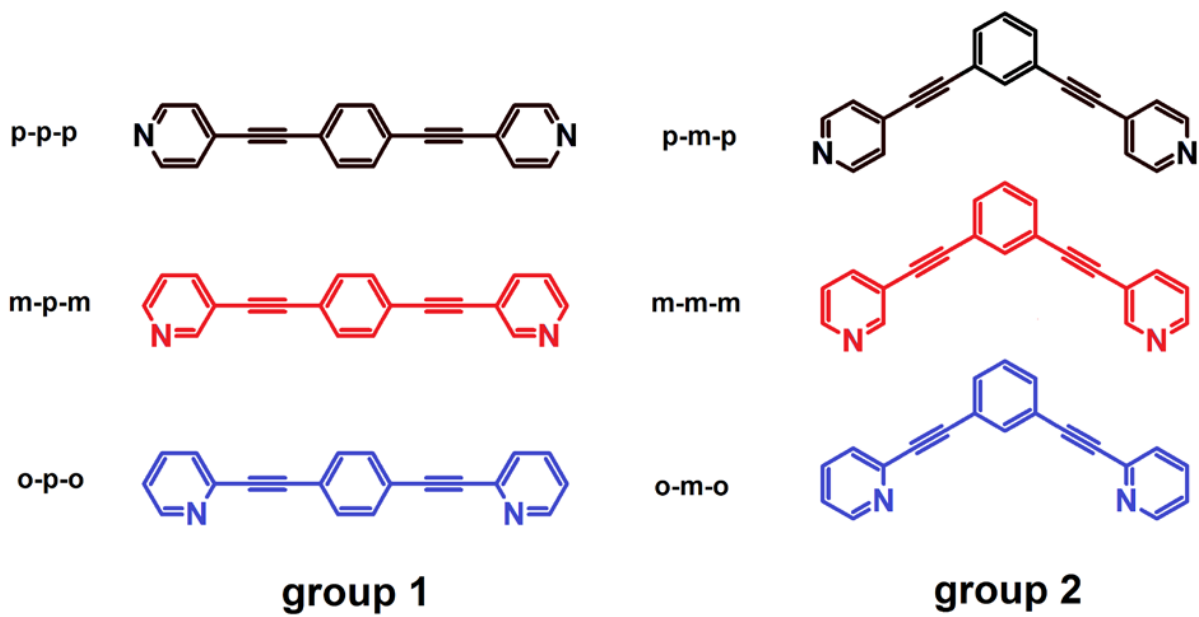

**Figure 1 | The molecular structures studied in this paper**. These are divided into two groups, based on the presence of a para (group 1) or meta (group 2) central phenyl ring.

RESULTS

**Break-junction experiments.** Charge transport characteristics of single-molecule junctions formed from the molecules in Figure 1 were investigated using both the mechanically-controllable-break-junction (MCBJ) and scanning-tunneling-microscopy-break-junction (STM-BJ) techniques, as reported elsewhere.[6,30,37,38] NMR spectra for these molecules and details of synthesis are presented in Supplementary Note 1 and Supplementary Figures 1-6 of the supplementary information (SI). Figure 2A displays typical traces of the conductance $G$ (in units of the conductance quantum $G_0=2e^2/h$) versus the relative electrode displacement ($\Delta z$) from measurement of the molecule p-p-p. Corresponding results for other molecules are presented in Supplementary Note 2, Supplementary Figures 7 to 12. $\Delta z$ is defined to be zero when $G=0.7 G_0$ and it is related to the electrode separation $z_{exp.}$ by $z_{exp.} = \Delta z + \Delta z_{corr}$, where the correction $\Delta z_{corr} =0.5 +/- 0.1$ nm accounts for the snap-back of the electrodes upon breaking of the gold-gold atomic contact[39]. For this molecule, the $\log(G/G_0)$ vs. $\Delta z$ stretching traces possess well-defined plateaus in the range of $\log(G/G_0)$ around -4.5, which we assign to the conductance of single-molecule junctions. The 2-dimensional (2D) histograms of p-p-p in Figure 2B show features of gold point contacts around $G \geq 1 \ G_0$ followed by a second accumulation in the cloud-like scatter plot in the range [$10^{-5.0} \ G_0 < G < 10^{-3.6} \ G_0$], centered at $G = 10^{-4.5} \ G_0$. We attribute the latter to the formation of single-molecule junctions. These clouds of conductance data lead to peaks in the corresponding 1-dimensional (1D) conductance histogram. The cloud-like pattern is observed in both MCBJ and STM-BJ measurements and the 1-D histogram peaks are in good agreement with each other. Figures 2C and 2D display the corresponding 1D conductance histograms of molecules belonging to groups 1 and 2 in a semi-logarithmic scale, constructed from 1000 experimental conductance-distance traces for each compound.

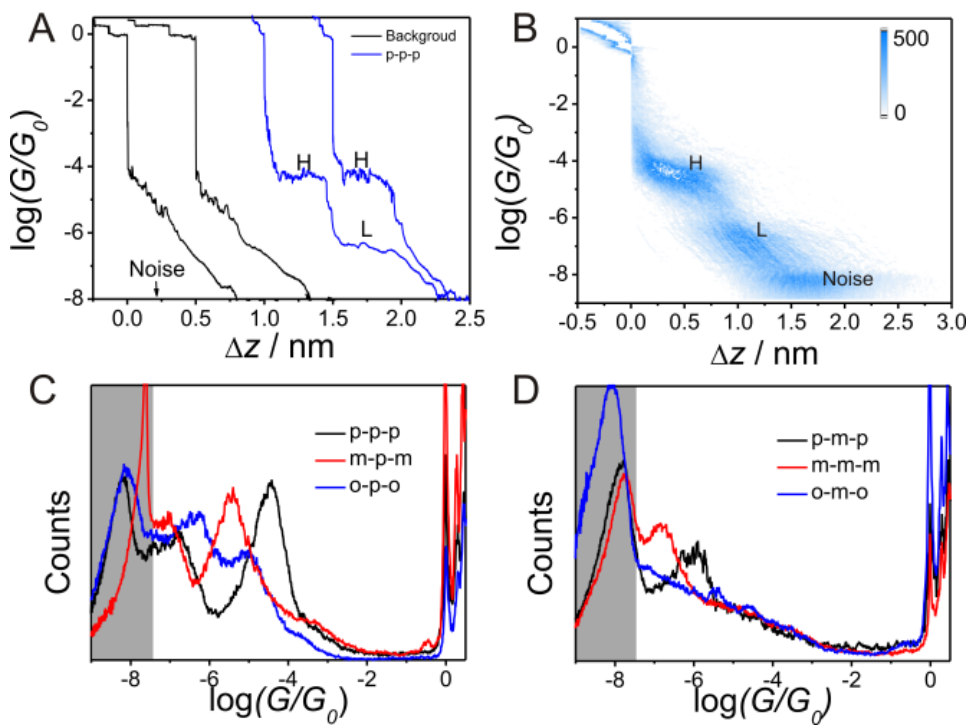

**Figure 2 | Conductance measurements.** (A) Typical individual conductance-distance traces of p-p-p (Blue) and pure tunneling traces (Black); (B) All-data-point 2D conductance vs relative distance ($\Delta z$) of p-p-p; (C, D) All-data-point 1D conductance histograms constructed from 1000 MCBJ traces of molecules in group 1 (C) and in group 2 (D). The grey area is the noise level.

The sharp peaks around 1 $G_0$ represent the conductance of a single-atom gold-gold contact. The prominent peaks between $10^{-7}$ $G_0$ $< G < 10^{-4}$ $G_0$ represent molecular conductance features. Significant difference of single-molecule conductances are observed from the variety of connectivities of the central ring (Figure 2C) and locations of the nitrogen in the anchor units (Figure 2D), while para connection in both central and terminal rings show the highest conductance in both cases. The statistically-most-probable conductance values were obtained by fitting Gaussians to the maxima in the conductance histograms. Our key results are summarized in Table 1. In anticipation of the theoretical discussion below, it is interesting to note that to within experimental error, $\log(G_{ppp}/G_0) + \log(G_{mmm}/G_0)$ (ie -4.5 - 6.9) is equal to $\log(G_{pmp}/G_0) + \log(G_{mpm}/G_0)$ (ie -5.5 - 6.0), which suggests that the product of the conductances for p-p-p and m-m-m molecules is equal to the product of conductances for p-m-p and m-p-m molecules and the quantum circuit rule $G_{ppp}/G_{pmp} = G_{mpm}/G_{mmm}$ is satisfied.

Further statistical analysis of conductance versus $\Delta z$ curves provides information about the junction formation probability (JFP) and allows us to determine the most probable relative electrode displacement ($\Delta z_H^*$) at the end of the high conductance plateaus. For every $log(G/G_0)$ vs. $\Delta z$ stretching trace, we determine the relative electrode displacement at the end of the high conductance plateau, $\Delta z_H$, which is the largest $\Delta z$ value within the range $-0.3 > log(G/G_0) > log(G_H^{end}/G_0)$, where $G_H^{end}$ is the end of high conductance feature. The most probable values of $\Delta z_H$ (denoted $\Delta z_H^*$) are obtained by constructing a histogram and fitting a Gaussian function to the largest maxima. Taking into account the snap-back length, the most probable electrode separations at the end of the high conductance plateau is $z_H^* = \Delta z_H^* + \Delta z_{corr}$. The representative $\Delta z_H$ histograms with Gaussian fitting functions are shown in Supplementary Figures 7-12. The junction formation probability (JFP) is calculated as the ratio of the area under the fitted Gaussian function and the total area of the $\Delta z_H$ histogram. If no distinct peak is observed in the $\Delta z_H$ histogram, then the junction formation probability is considered to be zero. Table 1 also summaries the various distances obtained from both MCJB and STM-BJ measurements. As shown in Table 1 the JFP approaches 100% for molecules p-p-p, p-m-p, m-p-m and m-m-m. For molecule o-p-o, the JFP decreased sharply to 21% (MCBJ)/27% (STM-BJ) because the N in the terminal ortho pyridyl is partially hidden from the electrode surfaces and therefore difficult to form a bridge between the two gold electrodes[40]. Our inability to measure the conductance of o-m-o is explained by its short N---N length and the expected low conductance, which falls below the direct tunneling conductance (Supplementary note 3 and Supplementary Figure 13).

**Table 1.** Most-probable experimental conductance, electrode separation $z_H^*$ at the end of the high-conductance plateaus and junction formation probability (JFP) of pyridyl terminated OPE derivatives from MCBJ and STM-BJ. The origin of differences between MCBJ and STM-BJ results are discussed in Supplementary Note 4. Error bars were determined from the standard derivation in Gaussian fitting of conductance and $\Delta z_H^*$ distribution. Comparison between theoretical lengths and most-probable end-of-plateau experimental electrode separations $z_H^*$. The electrode separation $z_H^*$ is closer to the theoretical N---N distance $L_{NN}$ than to the theoretical molecular length L.

| Molecule | Conductance (Log (G/G₀)) | | | JFP(%) | | $z_H^* = \Delta z_H^* + \Delta z_{corr}$ (nm) | | Theoretical Lengths | |
|---|---|---|---|---|---|---|---|---|---|
| | MCBJ | | STM-BJ | MCBJ | STM-BJ | MCBJ | STM-BJ | L (nm) | $L_{NN}$ (nm) |
| | High | Low | | | | | | | |
| p-p-p | -4.5±0.4 | -7.0±0.7 | -4.5±0.4 | 100 | 100 | 1.58±0.21 | 1.80±0.30 | 1.66 | 1.66 |
| m-p-m | -5.5±0.4 | -7.1±0.7 | -5.5±0.5 | 100 | 100 | 1.50±0.16 | 1.45±0.17 | 1.65 | 1.53 |
| o-p-o | -5.0±0.3 | -6.3±0.7 | -4.5±0.4 | 21 | 27 | 1.22±0.25 | 1.14±0.11 | 1.65 | 1.25 |
| p-m-p | -6.0±0.5 | - | -5.8±0.2 | 100 | 100 | 1.37±0.21 | 1.31±0.18 | 1.38 | 1.35 |
| m-m-m | -6.9±0.5 | - | <-6 | 100 | - | 1.36±0.14 | - | 1.40 | 1.40 |
| o-m-o | - | - | - | - | - | - | - | 1.40 | 1.17 |

The most-probable end-of-plateau electrode separations $z_H{}^*$ follow the trends p-p-p > m-p-m > o-p-o, and p-m-p ≈ m-m-m that correlate with the molecular N---N distance, demonstrating that the gold-anchor link is primarily controlled by the gold-nitrogen bonds. Therefore, it is clear that changes in the position of the N atom within the anchors affects both the plateau length and the JFP, as well as the conductance.

**Quantum interference in the terminal and central rings.** For the linear molecules of group 1, both p-p-p and o-p-o have the same conductance value ($10^{-4.5}$ $G_0$) from the STM-BJ technique. The conductance value of o-p-o from MCBJ with the Gaussian fitting is $10^{-5.0} G_0$, which is lower than that obtained using the STM-BJ. We also note that the width of the conductance peak in the 1D histogram from MCBJ measurements is bigger than that from the STM-BJ. This is associated with the different stretching of the molecular junctions during their formation using the two techniques, as verified by the 2D conductance properties (See SI). Nevertheless, compared to the STM-BJ results, the peak value from Gaussian fitting to MCBJ measurements is lower. These results suggest that a wider variety of configurations were formed between the molecule and electrodes during junction stretching in MCBJ measurements, because of the higher stability, while some of the molecular configurations, especially the fully-stretched configurations, could be monitored using the STM-BJ. The conductance of the m-p-m molecule was found to be lower than the other two group 1 members, with o-p-o and p-p-p in both the STM-BJ and MCBJ experiments, having a measured value of $10^{-5.5}$ $G_0$.

Comparisons between the conductances of molecules in group 2 also demonstrate that the para/meta/ortho variations in the terminal ring result in only small conductance changes of less than an order of magnitude. The measured conductance of m-m-m is $10^{-6.9}$ $G_0$, which is lower than the p-m-p conductance of $10^{-6.0}$ $G_0$. However as mentioned above, we were unable to measure any electrical properties of o-m-o from either MCBJ or STM-BJ, due to the masking direct tunneling current. For comparison, in ref 41, changing from para to meta in the central ring of an OPV3 caused the conductance to drop by one order of magnitude, while in ref. 32 changing the position of an -SMe or and -$NH_2$ in the anchor groups from para to meta caused a change of almost two orders of magnitude. These are different anchor groups from ours, but the conductance changes are comparable with our experimental observations.

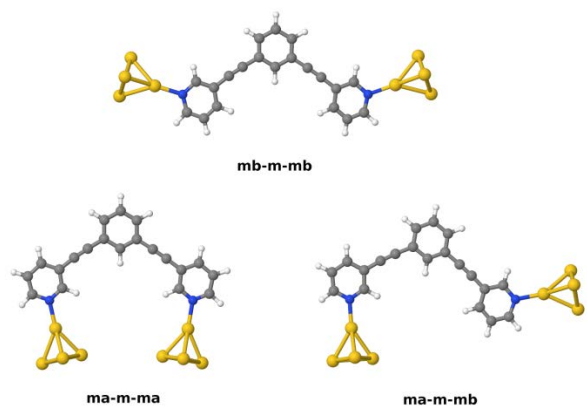

**Figure 3 | Molecular conformations**. Idealized junction geometries for the planar m-m-m molecule are shown. The N---N distance varies with conformations, therefore the molecular N---N distance, $L_{NN}$ is defined as the largest N---N distance (mb-m-mb junction geometry).

To elucidate the measured conductance trends we performed a density functional theory (DFT) based optimization of each molecule using SIESTA[42,43] and carried out electron transport calculations using the approach discussed in our previous papers[39,44]. Unlike the p-p-p molecule, the other molecules possess anchoring nitrogen atoms located at meta or ortho positions within the terminal rings, which do not naturally bind to planar electrode surfaces. To study the transport properties of these molecules, we attached the anchor nitrogen atoms to the apexes of pyramidal gold electrodes, as shown in Figure 3, Figure 4 and in Supplementary Note 5, Supplementary Figure 14. The molecules shown in these figures have been geometrically relaxed using SIESTA and it is clear that many of these geometries, such as the ma-m-ma conformation in Figure 3, would not tend to form a bridge between two opposing pyramids. We also note that molecular lengths ($L$), defined as the distance between the centers of the furthest non-hydrogen atoms in the molecule, are the same for these different conformations, whereas the N---N distance, which is the distance between the centers of the N atoms, does vary. Therefore, in Table 1 the largest N---N distance ($L_{NN}$) of each molecule is calculated and compared with the most probable electrode separation at the end of the high conductance plateau, i.e. $z_H^*$.

The central focus of this section is to understand the relative contribution to the conductance from QI in the terminal and central aromatic rings and to demonstrate that the quantum circuit rule is satisfied at the level of DFT. The largest value of $z_H^*$ amongst all the measured molecules (Table 1) was found to be $z_H^* = 1.80$ nm for the p-p-p. The value of $z_H^*$ for other molecules was typically shorter than this value. Therefore the typical conductances occur when the molecules are not fully stretched and there is a possibility of metal-ring overlap between the gold electrode and the pyridyl rings. To analyze such events we systematically constructed junction geometries with two 35 atom gold (111) directed pyramids attached to the N atoms perpendicular to the pyridyl rings, shown in Figure 4 and then computed the transmission coefficient curves $T(E)$ for these geometries. (Alternative junction geometries are investigated in the SI, Supplementary Note 6, Supplementary Figures 15, 16 and 17. Supplementary Figure 18 shows a comparison between theoretical and experimental conductances.)

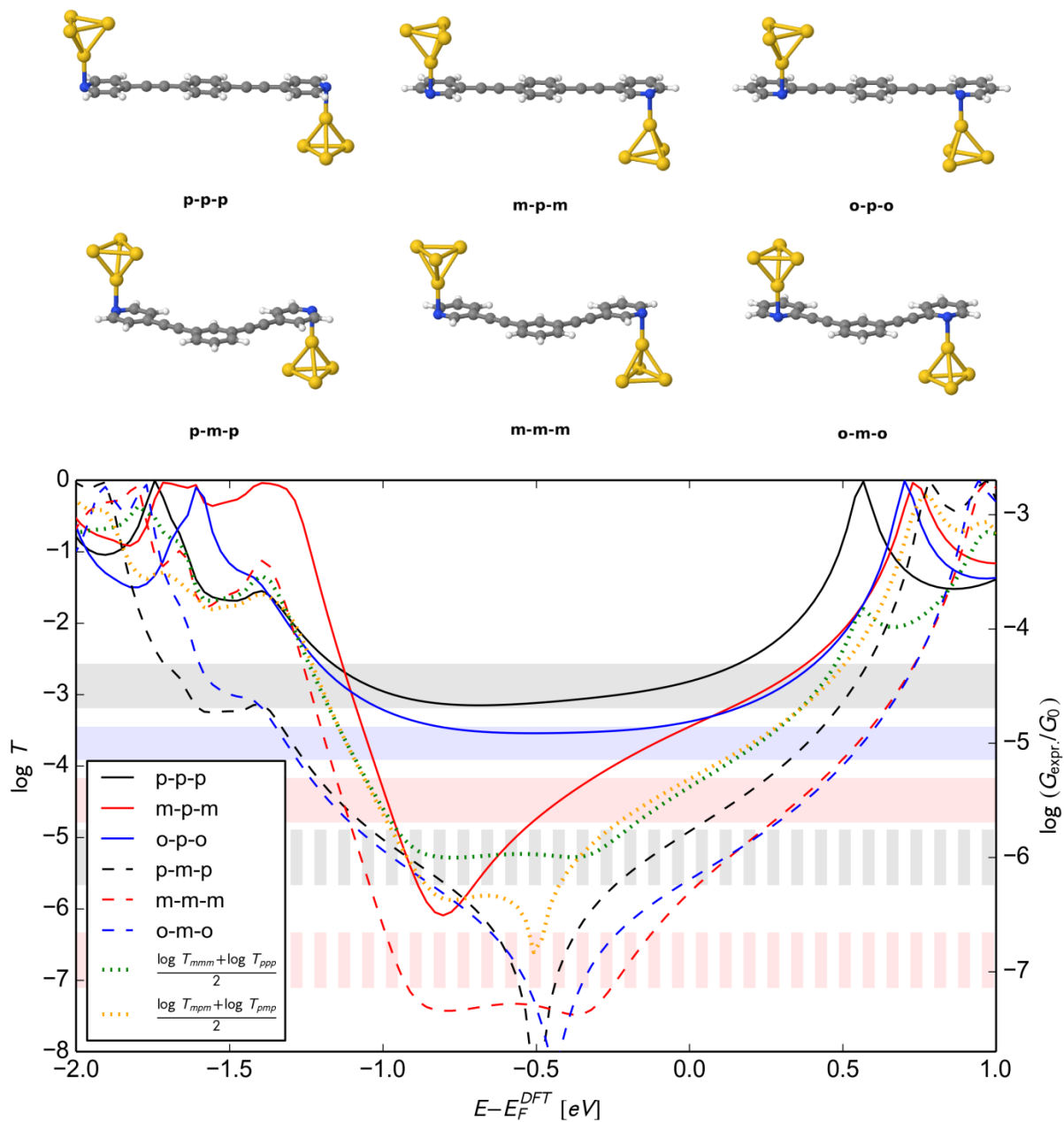

**Figure 4 | Systematic junction geometries and transmission coefficient functions.** The top figures show idealized junction geometries for all six molecules with the gold tip attached perpendicular to the pyridine ring. The bottom graphs show corresponding transmission coefficient curves. For comparison, the colored and patterned horizontal shaded bars show the experimentally measured MCBJ values of log ($G/G_0$) (right hand scale). The thickness of the bars corresponds to the error bar of the experimental log ($G/G_0$). These show that both the transmission coefficients around the $E_F$ and the experimental conductance values are separable into two groups corresponding to X-p-X and X-m-X connectivities. The green (yellow) dotted line is the average of log $T_{mmm}$ and log $T_{ppp}$ (log $T_{mpm}$ and log $T_{pmp}$). It should be noted that over most of the energy range -0.8 to -0.4 eV, the m-m-m results have the lowest transmission coefficient. At these low transmissions the contribution from the σ channel becomes comparable with that of the π channel and therefore in this energy range, the circuit rule is violated.

Figure 4 demonstrates that for wide range of Fermi energy choice the theoretical and experimental conductances of the X-p-X molecules of group 1 are distinctly higher than those of the molecules in group 2, as expected from previous studies[27,32,41,45]. Figure 4 also shows that for wide range of energies in the HOMO-LUMO gap, the ordering of the transmission coefficients follows the experimental conductance ordering. To demonstrate the resilience of the circuit rule, dotted lines are plots of (log $T_{mmm}$ + log $T_{ppp}$ )/2 and (log $T_{mpm}$ + log $T_{pmp}$ )/2. The similarity of these two curves shows that the product rule is satisfied over a wide range of energies within the HOMO-LUMO gap.

To demonstrate that QI effects associated with variations in the positions of the N atoms are suppressed due to the presence of a parallel conductance path associated with the electrode-ring overlap, we performed transport calculations in which we artificially set to zero every Hamiltonian and overlap matrix element that couples carbon and hydrogen atoms to gold atoms, leaving couplings between the nitrogen atoms and gold as the only possible transport path. The resulting transmission coefficients are shown as dashed lines in Figure 5 and demonstrate that without metal-ring interactions the meta link in the terminal ring of the m-p-m molecule reduces the conductance by orders of magnitude, which is comparable with the effect of a meta link in the central ring, whereas in the presence of metal-ring interactions, meta-coupling in the terminal rings has a much smaller effect.

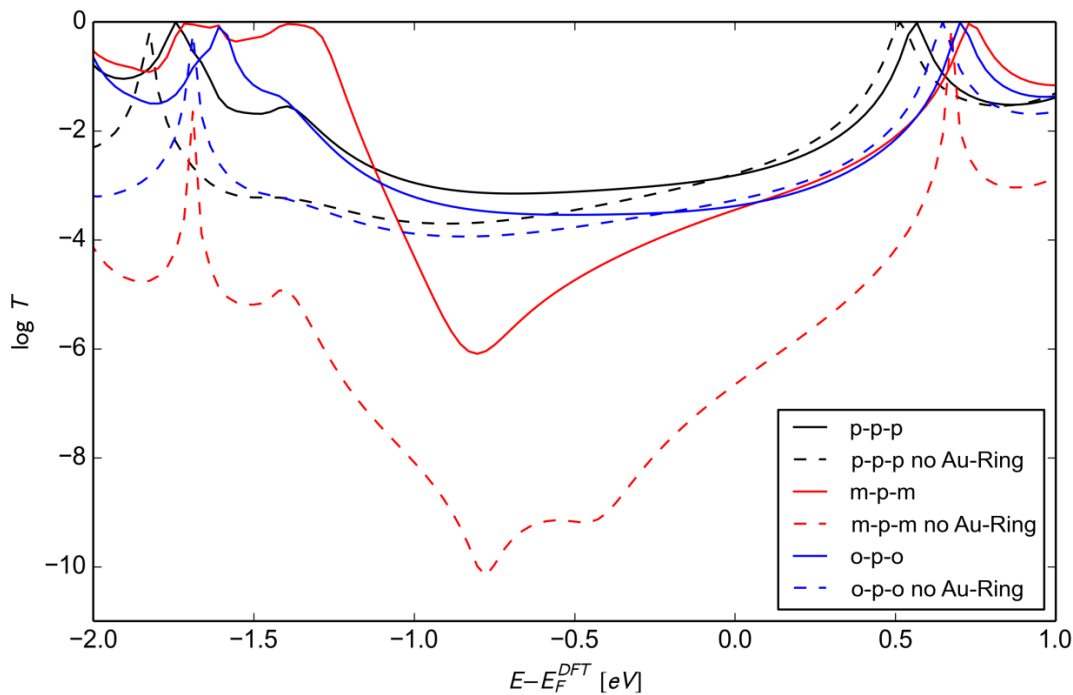

**Figure 5 | Transmission curves with and without metal-ring coupling.** The corresponding junction geometries are shown in Figure 4. Dashed lines are without metal-ring coupling, the continuous curves are with metal-ring coupling.

**Derivation of the quantum circuit rule for molecular conductances.** Having demonstrated that the effect of QI in a given ring depends on the position of the ring within the molecule (ie whether it is a terminal or a central ring), we now derive the following quantum circuit rule describing the conductances of different ring combinations:

$$G_{ppp} \times G_{mmm} = G_{pmp} \times G_{mpm} \quad (1)$$

This captures the property that the contribution to the conductance from the ring Y in molecules of type p-Y-p and m-Y-m are identical.

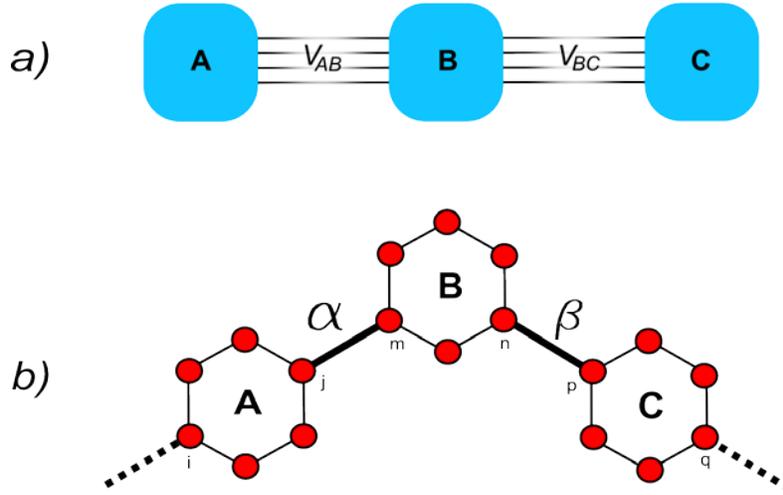

**Figure 6 | Abstract models.** (a) An abstract depiction of three coupled quantum sub-systems. (b) A simple tight-binding model of the three ring system. (In the figure p-m-p connection is shown to illustrate the connecting sites i,j,m,n,p and q.)

To derive the circuit rule, consider the simple tight-binding model of Figure 6b, in which three rings (labelled A, B, C) are coupled by nearest neighbor couplings α and β. To calculate the transmission coefficient arising when 1-dimensional electrodes are connected to site i of ring A and q of ring C, we first calculate the Green's function $G_{qi}$ of the whole structure in the absence of the electrodes.

This structure is an example of three arbitrary coupled quantum objects shown schematically in Figure 6a, whose Green's function is given by

$$\begin{bmatrix} E - H_A & -V_{AB} & 0 \\ -V_{AB}^\dagger & E - H_B & -V_{BC} \\ 0 & -V_{BC}^\dagger & E - H_C \end{bmatrix} \begin{bmatrix} G_{AA} & G_{AB} & G_{AC} \\ G_{BA} & G_{BB} & G_{BC} \\ G_{CA} & G_{CB} & G_{CC} \end{bmatrix} = I.$$

Electron propagation from sub-system A to sub-system C is described by the block $G_{CA}$, for which the above equation yields

$$G_{CA} = g_C V_C^\dagger G_{BB} V_A^\dagger g_A,$$

where $g_X = (E - H_X)^{-1}$ (X=A or C) and $G_{BB} = \left(E - H_B - V_{AB}^\dagger g_A V_{AB} - V_{BC} g_C V_{BC}^\dagger\right)^{-1}$. As an example, for the tight binding model of Figure 6b, this becomes $G_{qi} = (g_C)_{qp} B_{nm} (g_A)_{ji}$, where $B_{nm} = \beta \frac{(g_B)_{nm}}{(1-\sigma)} \alpha$ and $\sigma = \alpha^2 (g_A)_{jj}(g_B)_{mm} + \beta^2 (g_C)_{pp}(g_B)_{nn} + \alpha^2\beta^2 (g_A)_{jj}(g_C)_{pp}[(g_B)_{nm}(g_B)_{mn} - (g_B)_{nn}(g_B)_{mm}]$.

The crucial point is that $B_{nm}$ does not depend on the choice of *p, q, i, j*, because due to rotational symmetry, $(g_A)_{jj}$ and $(g_C)_{pp}$ are independent of *j* and *p* for any choice of *j* and *p*. This means that $B_{nm}$ does not depend on the choice of meta, ortho or para coupling for rings A and C and only depends on the connectivity of ring B. When the rings A and C are coupled to the electrodes, this rotational symmetry is broken slightly. However, provided $E_F$ lies in the HOMO-LUMO gap of A and C and the coupling to the electrodes is small, (as is usually the case for molecules attached to gold electrodes, where the level broadening is much less than the level spacing) then the symmetry breaking is weak[26]. Since the electrical conductance is proportional to $|G_{qi}|^2$ this means that the electrical conductance $G_{XYZ}/G_0$ of molecules of the type X-Y-Z is proportional to a product of the form $a_X \times b_Y \times a_Z$ and hence the quantum circuit rule is satisfied. The factors $a_X, b_Y, a_Z$ are contributions to the overall conductance from the separate rings, but it should be noted that they are not themselves individually-measureable conductances. The above analysis can be easily applied to a molecule with two rings yield the circuit rule $G_{pp}G_{mm} = G_{pm}^2$, where the $G_{pp}$ and $G_{mm}$ are the conductances of the molecule with para-para and meta-meta connected rings, and $G_{pm}$ is the conductance of the molecule with meta and para rings (see SI for a DFT confirmation of this rule).

To check the experimental validity of the circuit rule (1) for the OPEs of Figure 1, we examine the measured conductance values presented in Table 1. For example, rearranging equation (1) into the form $G_{pmp} = \frac{G_{ppp}}{G_{mpm}} G_{mmm}$ and substituting into the right hand side the measured values for $G_{ppp}, G_{mmm}$ and $G_{mpm}$ yields $G_{pmp} = 10^{-5.9} G_0$, which compares well with the measured value of $G_{pmp} = 10^{-5.8} G_0$ shown in table 1.

DISCUSSION

We have presented charge transport studies of pyridyl terminated OPE derivatives, using the MCBJ and STM-BJ techniques, DFT-based theory and analytic Green's functions and have investigated the interplay between QI effects associated with central and terminal rings in molecules of the type X-Y-X.

Our results demonstrated that the contribution to the conductance from the central ring is independent of the para or meta nature of the anchor groups and the combined conductances satisfy the quantum circuit rule $G_{ppp}/G_{pmp} = G_{mpm}/G_{mmm}$. For the simpler case of a two-ring molecule, the circuit rule $G_{pp}G_{mm} = G_{pm}^2$ is satisfied (Supplementary Note 8, Supplementary Figures 19 and 20). It should be noted that the circuit rule does not imply that the conductance $G_{XYX}$ is a product of three measureable conductances associated with rings X, Y and X. Indeed the latter property does not hold for a single molecule. On the other hand provided sample to sample fluctuations lead to a broad distribution of phases within an ensemble of measurements a product rule for ensemble averages of conductances can arise. This possibility is discussed in detail in Supplementary Note 9.

The qualitative relationship between the conductances agrees well with the simple quantum interference picture of molecular conduction. It has been reported that destructive QI exists in benzene with the meta connectivity and is responsible for the observed reduction of conductance[16,20,28], whereas for para and ortho connectivities, constructive QI should be observed[46,47]. The transmission coefficient calculations through junctions where the metal-ring connection is artificially blocked (Figure 5) show that the artificially-coupled pyridyl ring exhibits similar behaviour to the benzene ring, with destructive QI in the case of the meta coupling significantly reducing the conductance compared with para and ortho connectivities. The dashed curves in the bottom panel in Figure 5 clearly demonstrate that when the conduction is through only the nitrogen atoms, the conductance of the meta isomer is much lower than in the para and ortho isomers. More realistically however, in the presence of metal-ring overlap, the effect of varying the positions of the nitrogens in the anchors becomes much weaker and as demonstrated by Figure 4, the major changes in the molecular conductance are caused by the variations in the connectivity of the central ring. The dominant influence of the central ring is accounted for by the fact that the central ring is not in direct contact with electrodes and therefore no parallel conductance paths are present, which could bypass the ethynylene connections to the anchors.

In a sub-nanometer scale molecular circuit, as in standard CMOS circuitry, electrical insulation is of crucial importance. Destructive interference in a 2-terminal device may not be desirable, because of the lower conductance. However for a three-terminal device minimizing the conductance of the third terminal is highly desirable, because the third (gate) electrode should be placed as close to the molecule as possible, but at the same time, there should be no leakage current between the molecule and gate. One way of achieving this may be to use an anchor group with built in destructive interference. Therefore destructive QI may be a vital ingredient in the design of future 3-terminal molecular devices and more complicated networks of interference-controlled molecular units.

METHODS

**Conductance measurements.** The transport characteristics in single-molecule junctions were studied by MCBJ and STM-BJ measurements in solution at room temperature. The latter contained typically 0.1 mM of the OPE-type molecules in a mixture of 1,3,5-dimethylbenzene (TMB; Aldrich, p.a.) and tetrahydrofuran (THF; Aldrich, p.a), 4:1 (v/v).

**Mechanically controlled break junction (MCBJ) measurements.** The MCBJ experiments are based on the formation and breaking of a nanogap between a notched, freely suspended gold wire (0.1 mm diameter, 99.999%, Goodfellow), fixed on spring steel sheets (10 mm × 30 mm, thickness 0.25 mm) with a two-component epoxy glue (Stycast 2850 FT with catalyst 9). The sample sheets were fixed between two holders. A Kel-F liquid cell with a Kalrez O-ring was mounted onto the sheet. During the measurements, the steel sheet could be bent with a pushing rod, which was controlled by a combination of a stepper motor and a piezo stack. The bending was initialized by the stepper motor. Once the measured current decreased to a value corresponding to 15 $G_0$ the stepper motor stops to move, and the piezo stack was activated. This strategy reduced significantly noise contributions from the operation of the stepper motor. The movement of the piezo stack controlled the breaking and the reformation of nanoscale contacts, typically in the range between the noise threshold ($G< 10^{-8}G_0$) and a high conductance limit, which was set to 10 $G_0$. Molecular junctions could form upon breaking the gold-gold nanocontacts. The entire cycle was repeated more than 1000 times to obtain statistically relevant data. In the MCBJ setup, the current could be recorded as a feedback signal at a given bias voltage (typically between 0.020 and 0.200 V). The two ends of the "broken wire" were taken as working electrodes WE1 and WE2.

The MCBJ unit is controlled by a lab-built bipotentiostat with two bipolar tunable logarithmic $I - V$ converters as current measuring units, which are operated by a custom-designed micro-controller. The system provides three analog signals: the potential of WE1, the voltage difference between the two working electrodes WE1 and WE2 (bias voltage $V_{bias}$), driving the current through the two gold electrodes for the conductance measurements, and the voltage output of the piezo stack in the range of 0 to 50 V, allowing the displacement of the piezo stack up to 8μm with rates ranging from 3 to 3000 nm s$^{-1}$. The latter translates into lateral pulling (pushing) rates between the two gold leads of 0.1 to 100 nm s$^{-1}$. The distance between the two gold electrodes in the MCBJ setup was calibrated with complementary STM-BJ experiments assuming that the tunneling decay is identical under the same experimental conditions. Further technical details of the MCBJ set-up reported by Hong et al. in ref[48].

**STM-BJ.** The STM-BJ technique repeatedly traps molecules between a sharp STM tip and an atomically flat sample.

The STM-BJ measurements were carried out with a Molecular Imaging PicoSPM housed in an all-glass argon-filled chamber and equipped with a dual preamplifier capable of recording currents in a wide range of 1 pA to 150 μA with high resolution. The non-amplified low-current signal was fed back to the STM controller preserving the STM imaging capability. The current-distance measurements were performed with a separate, lab-build analog ramp unit. For further technical details we refer to our previous work[49,50].

The sample electrode was gold single crystal bead. The Au (111) facet prior to each experiment was subjected to electrochemical polishing and annealing in a hydrogen flame followed by cooling under Ar. A freshly prepared solution containing typically 0.1 mM of the respective molecule was added to a Kel-F flow-through liquid cell mounted on top of the sample and the uncoated STM tip was electrochemically etched using gold wire (Goodfellow, 99.999%, 0.25 mm diameter) capable of imaging with atomic resolution. This system relies on trapping a molecule between the end of an Au tip and the gold substrate.

After assembling the experiment, the following protocols were applied: The tip was brought to a preset tunneling position typically defined by $i_T$ = 50 to 100 pA and a bias voltage $V_{bias}$ = 0.10, 0.065, and 0.175 V, followed by imaging the substrate.

After fixing the lateral position of the tip, the STM feedback is switched off and current-distance measurements were performed and then the vertical movement of the tip controlled by the ramp unit. The measuring cycle was performed in the following way: The controlling software drives the tip towards the adsorbate-modified surface. The approach was stopped until a predefined upper current limit was reached (typically 10 μA or < 10 $G_0$ with $G_0$ being the fundamental conductance quantum 77.5 μS). After a short delay (~100 ms) ensuring tip relaxation and the formation of stable contacts, the tip was retracted by 2 to 5 nm until a low current limit of ~10 pA was reached. The approaching and withdrawing rates were varied from 56 to145 nm/s. The entire current-distance traces were recorded with a digital oscilloscope (Yokogawa DL 750, 16 bit, 1 MHz sampling frequency) in blocks of 186 individual traces. Up to 2000 traces were recorded for each set of experimental conditions to guarantee the statistical significance of the results. For each molecule the data were acquired at three different bias voltages of 0.065, 0.100 and 0.175 bias voltage.

**Theory and simulations.** All molecules in this study were first geometrically optimized in isolation. For molecules with meta and ortho terminal pyridyl rings, all the non-equivalent planar conformations were geometrically relaxed. All geometrical optimizations were carried out using the DFT code SIESTA, with a generalized gradient approximation method (PBE functional), double-zeta polarized (DZP) basis set, 0.01 eV/ Å force tolerance and 250 Ry mesh cut off. To investigate ideal junction geometries, a small 4-atom gold pyramid was attached to the N atoms of the molecules, with Au-N-C angle being 120$^o$ and Au-N bond length 2.1 Å, as shown in Supplementary Figure 14. Many of these junction geometries are unlikely to happen in BJ experiments and during the junction elongation typically the gap between the electrodes is shorter than the molecular length. We constructed another set of idealized junction geometries where the gold pyramid is attached to rings from the side, perpendicular to the ring. These junction geometries are shown in Figure 4. For transport calculation the 4-atom gold pyramids that are seen in Figure 4 are extended to a 35-atom gold pyramid. Tranmission coefficients were computed using the GOLLUM code[51].

In break-junction experiments more complicated structures are expected. For this reason we also carried out BJ simulations using the approach discussed in our previous papers[39,44]. The geometrically-optimized molecules were inserted between two opposing 35 atom (111) directed pyramids with four different tip separations (The tip separation is defined to be the center to center distance between the apex atoms of the two opposing pyramids). In the initial geometries, the molecules were shifted slightly towards one of the pyramids and the initial Au-N distances were around 2.5 Å. Then the constructed structure was geometrically relaxed such that the base layers of the pyramids were kept fixed during the optimizations. Optimized junction geometries are shown in Supplementary Figures 16-17.

To produce conductance-trace curves in Supplementary Figure 18, the transmission coefficient *T(E)* was calculated for each relaxed junction geometry, by first obtaining the corresponding Hamiltonian and overlap matrices with SIESTA and DZP basis set. The transmission coefficient for all junction geometry in this study was calculated using wide-band electrodes with Γ = 4.0 eV . The wide-band electrodes were coupled to the two base layers of gold atoms of the 35-atom pyramids. A few additional transmission coefficient functions can be seen in Supplementary Figure 15. To produce conductance-trace curves, the transmission coefficient $T(E)$ was calculated for each relaxed junction geometry and the conductance $G/G_0 = T(E_F)$ was obtained by evaluating $T(E)$ at the Fermi energy $E_F$. A value of $E_F$ = -0.65 eV, relative to the bare DFT Fermi energy ($E_F^{DFT}$) was chosen to account for the expected LUMO level shift, as discussed in ref[39], where the Fermi energy shift was adjusted to match measured and computed conductance for the pyridine terminated oligoynes molecules.

**Synthesis**. Details of the synthesis are in the Supplementary Note 1.

**ACKNOWLEDGMENTS**

This work was generously supported by the Swiss National Science Foundation (200020-144471; NFP 62), the German Science Foundation (priority program SPP 1243), the UK EPSRC grants EP/K001507/1, EP/J014753/1, EP/H035818/1, the EC FP7 ITN "MOLESCO" project number 606728, and the University of Bern.

**Author Contributions**

All authors provided essential contributions to the manuscript and the project. All authors have given approval to the final version of the manuscript.

Additional information

Experimental part: synthesis and characterization of compounds, NMR spectra, transport measurements and data analysis procedures. Theoretical part: analytic details, computational details, ideal junction geometries, junction geometries for simulated BJ process, transmission curves.

**Corresponding Author**

c.lambert@lancaster.ac.uk; m.r.bryce@durham.ac.uk; hong@dcb.unibe.ch

**Notes**

The authors declare no competing financial interest. C. L., D.M., T. W., W. H., and M.R.B. originally conceived the concept and designed the experiments. D. M., C. H, W. H, M. R. B., and C. L. prepared the manuscript using feedback from other authors. Synthetic work was carried out in M. R. B.'s lab by X. Z. and M. G.; break junction measurements were carried out in T. W.'s lab by C. H., M. B, and V. K.; calculation were carried out in C. L.'s lab by D. M., O. A., and H. S.

Supplementary Figures

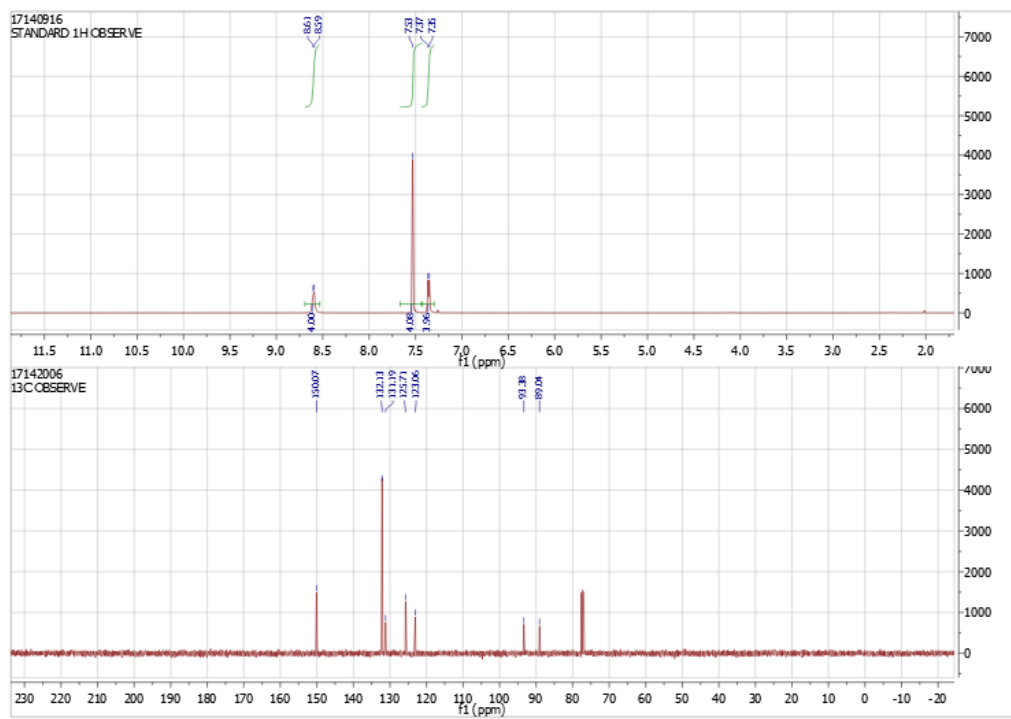

**Supplementary Figure 1** $^1$H and $^{13}$C NMR of **p-p-p** in CDCl$_3$

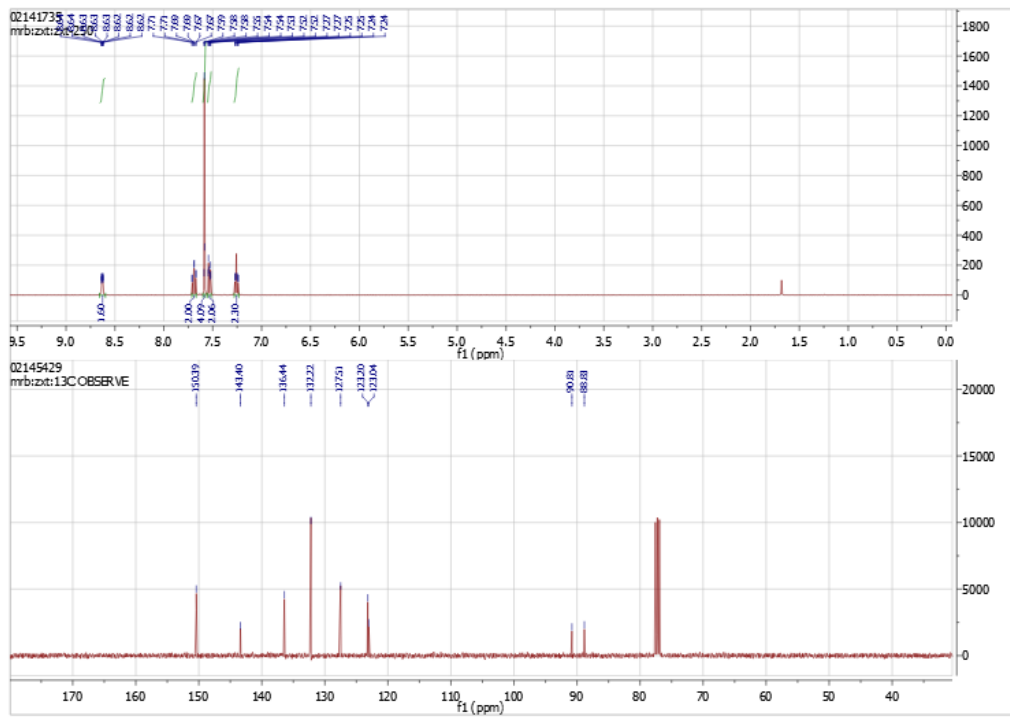

**Supplementary Figure 2** $^1$H and $^{13}$C NMR of **o-p-o** in CDCl$_3$

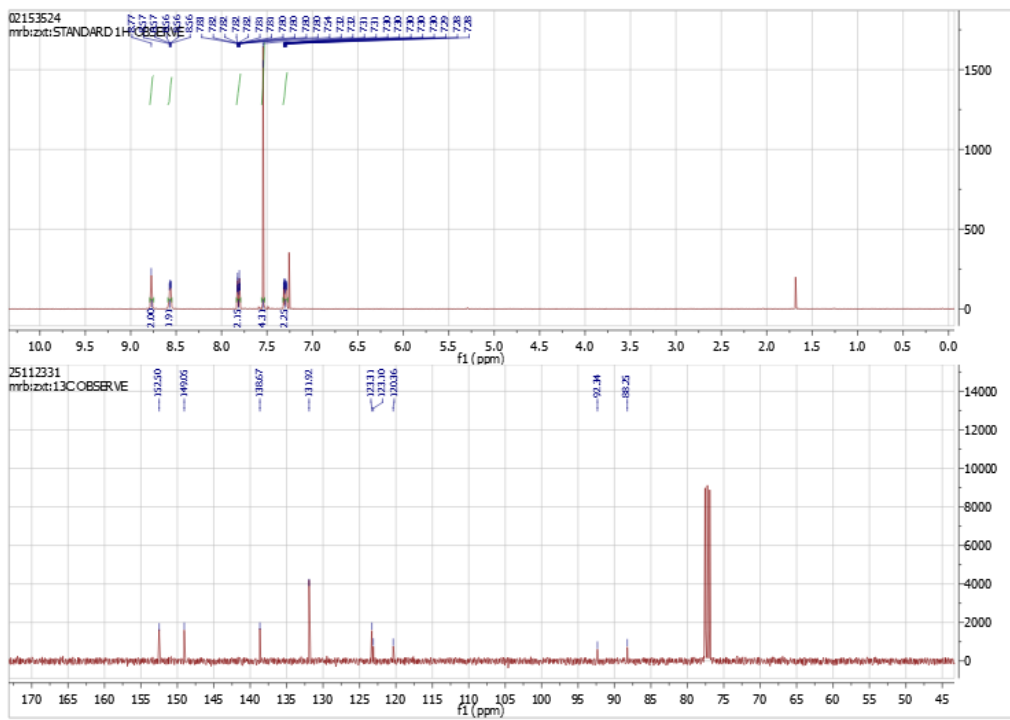

**Supplementary Figure 3** $^1$H and $^{13}$C NMR of **m-p-m** in CDCl$_3$

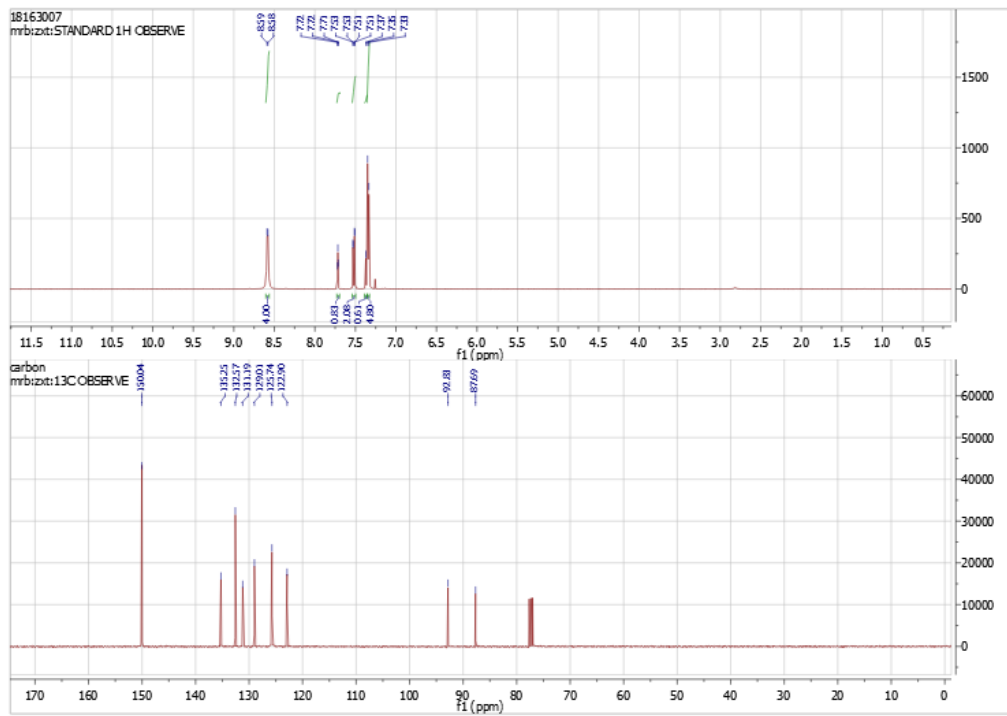

**Supplementary Figure 4** $^1$H and $^{13}$C NMR of **p-m-p** in CDCl$_3$

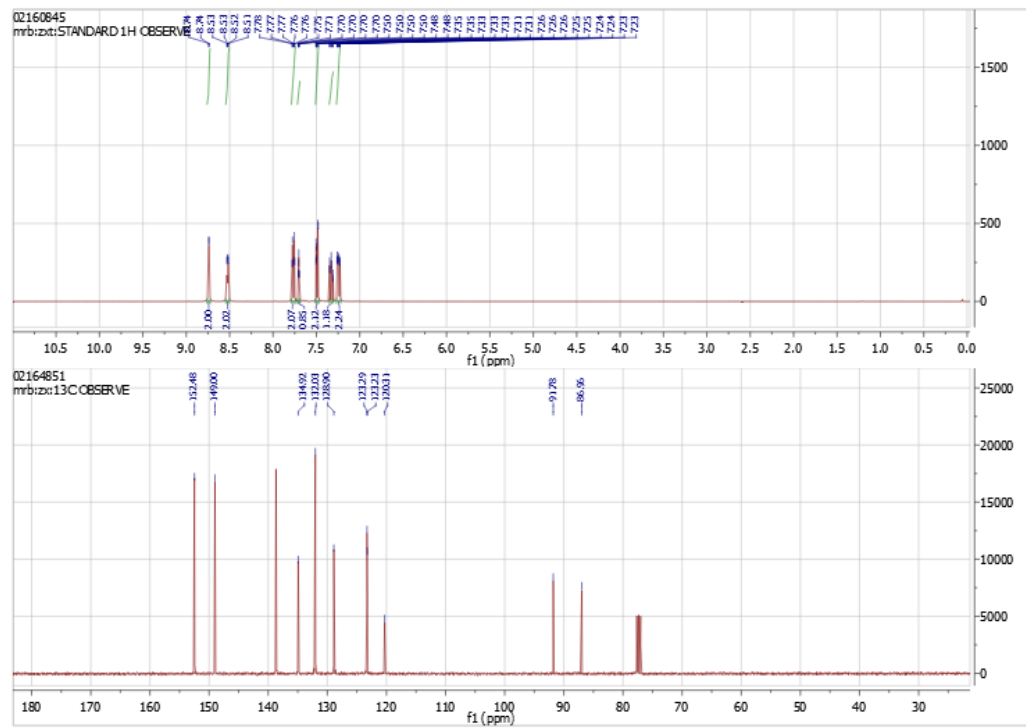

**Supplementary Figure 5** $^1$H and $^{13}$C NMR of **m-m-m** in CDCl$_3$

**Supplementary Figure 6** $^1$H and $^{13}$C NMR of **o-m-o** in CDCl$_3$

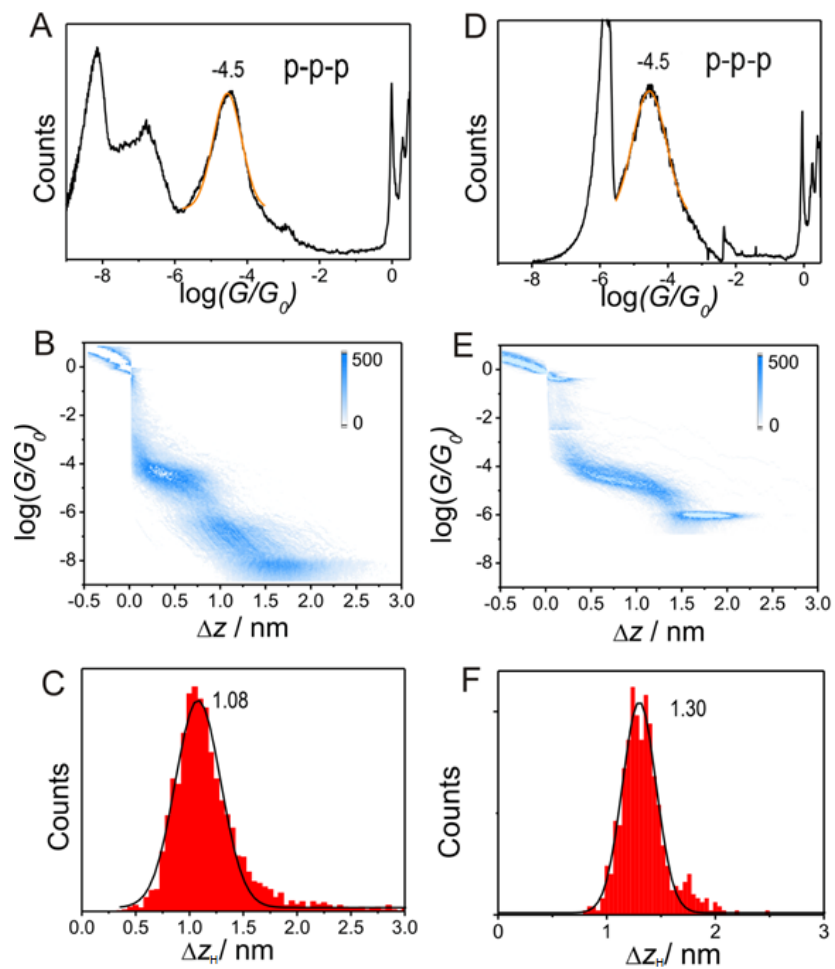

**Supplementary Figure 7.** 1D conductance histograms (A, D), 2D histograms (B, E), and distributions of the 'end-of-high-conductance-plateau displacements' $\mathit{\Delta z_H}$ (C, F) of p-p-p. The peaks of the latter are identified with $\mathit{\Delta z_H}^*$. Columns (A-C) were obtained using MCBJ and (D-F) using STM-BJ. $G_H^{end}$ chosen to construct the $\mathit{\Delta z_H}$ distribution is $10^{-5.8}\,G_0$ for MCBJ and $10^{-5.6}\,G_0$ for STM-BJ.

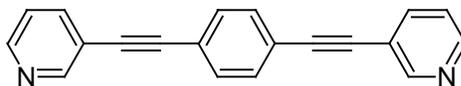
23x

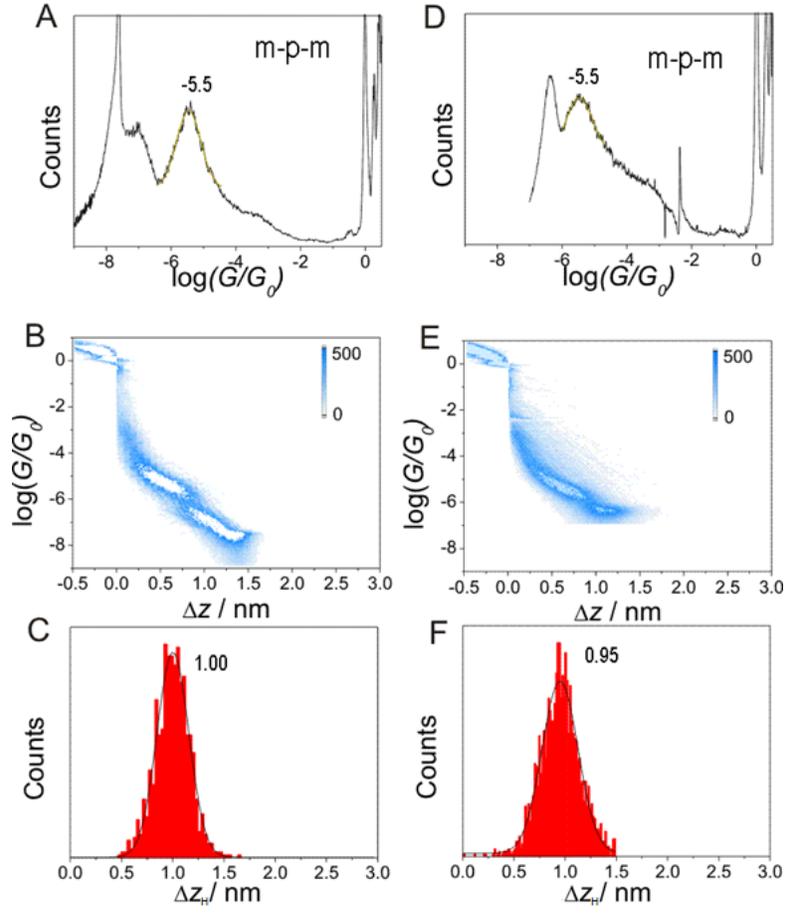

**Supplementary Figure 8.** 1D conductance histograms (A, D), 2D histograms (B, E), and distributions of the 'end-of-high-conductance-plateau displacements' $\Delta z_H$ (C, F) of m-p-m. The peaks of the latter are identified with $\Delta z_H^*$. Columns (A-C) were obtained using MCBJ and (D-F) using STM-BJ. $G_H^{end}$ chosen to construct the $\Delta z_H$ distribution is $10^{-6.4}\,G_0$ for MCBJ and $10^{-6.0}\,G_0$ for STM-BJ.

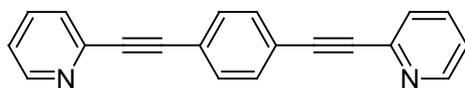

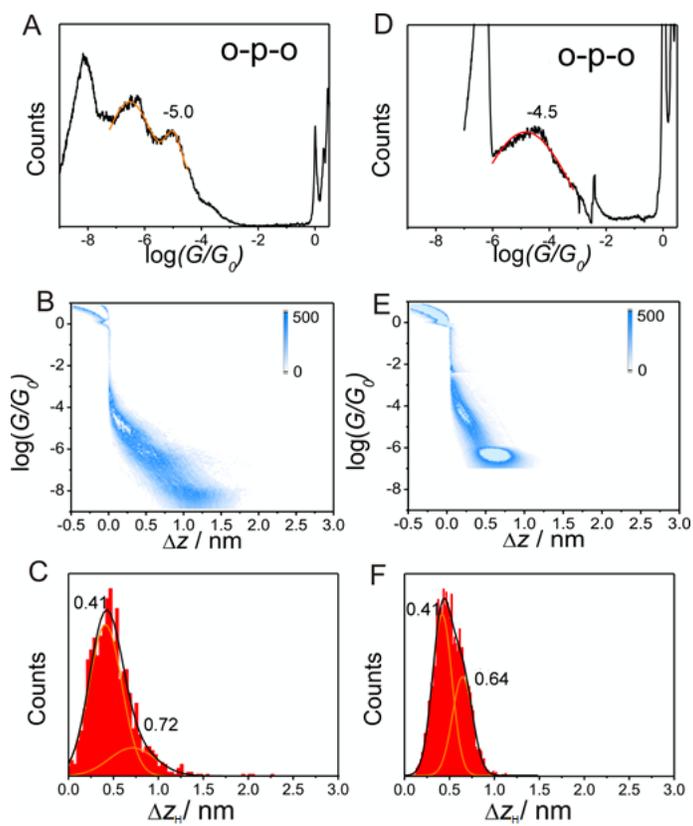

**Supplementary Figure 9.** 1D conductance histograms (A, D), 2D histograms (B, E), and distributions of the 'end-of-high-conductance-plateau displacements' $\Delta z_H$ (C, F) of o-p-o. The peaks of the higher- $\Delta z_H$ Gaussian fits are identified with $\Delta z_H^*$. Columns (A-C) were obtained using MCBJ and (D-F) using STM-BJ. $G_H^{end}$ chosen to construct the $\Delta z_H$ distribution is $10^{-5.6}\,G_0$ for MCBJ and $10^{-5.9}\,G_0$ for STM-BJ.

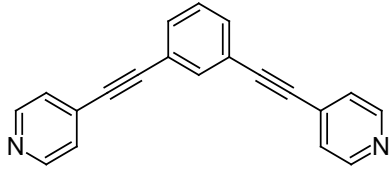

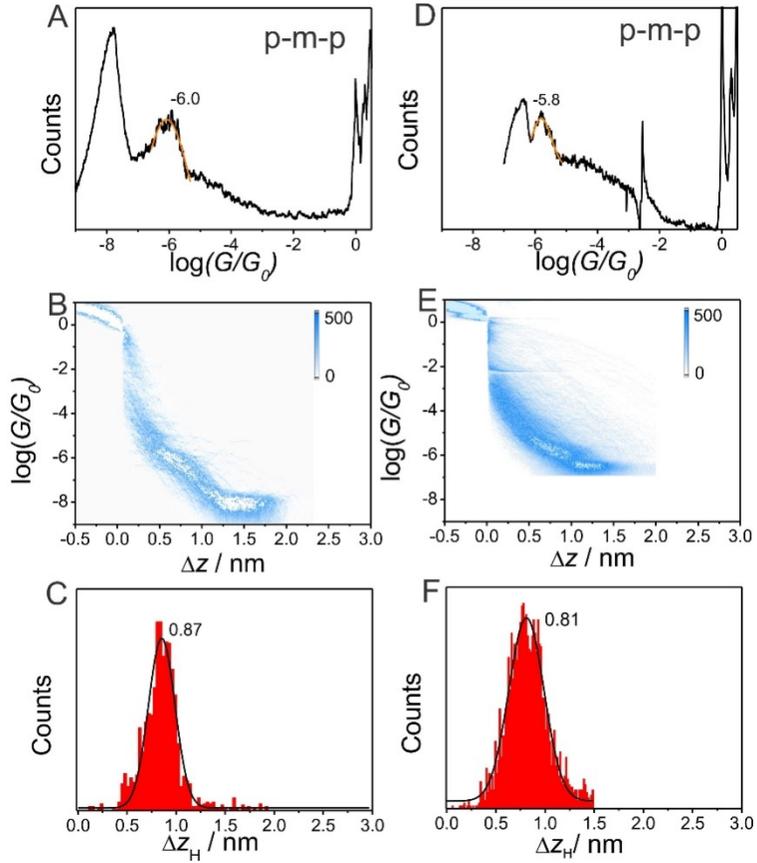

**Supplementary Figure 10.** 1D conductance histograms (A, D), 2D histograms (B, E), and distributions of the 'end-of-high-conductance-plateau displacements' $\Delta z_H$ (C, F) of p-m-p. The peaks of the latter are identified with $\Delta z_H^*$. Columns (A-C) were obtained using MCBJ and (D-F) using STM-BJ. $G_H^{end}$ chosen to construct the $\Delta z_H$ distribution is $10^{-7.5} G_0$ for MCBJ and $10^{-6.1} G_0$ for STM-BJ.

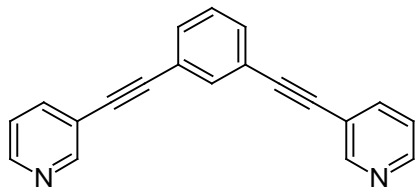

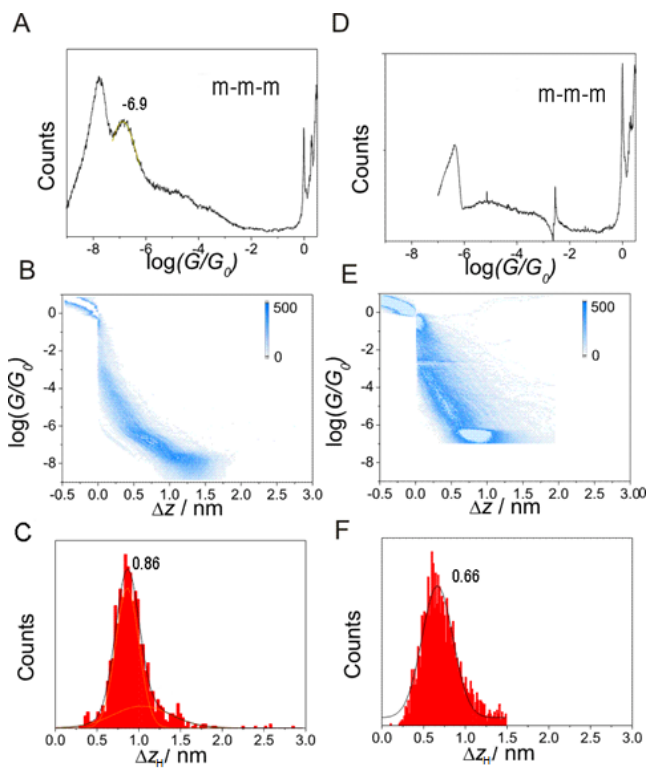

**Supplementary Figure 11.** 1D conductance histograms (A, D), 2D histograms (B, E), and distributions of the 'end-of-high-conductance-plateau displacements' $\varDelta z_H$ (C, F) of m-m-m. The peak of C is identified with $\varDelta z_H^*$. Columns (A-C) were obtained using MCBJ and (D-F) using STM-BJ. For this molecule, conductance measurement from STM provided pure tunneling traces. Therefore, the peak shown in F was assigned to the pure tunneling. $G_H^{end}$ chosen to construct the $\varDelta z_H$ distribution is $10^{-7.2}\,G_0$ for MCBJ and $10^{-6.0}\,G_0$ for STM-BJ.

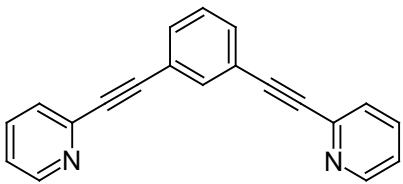

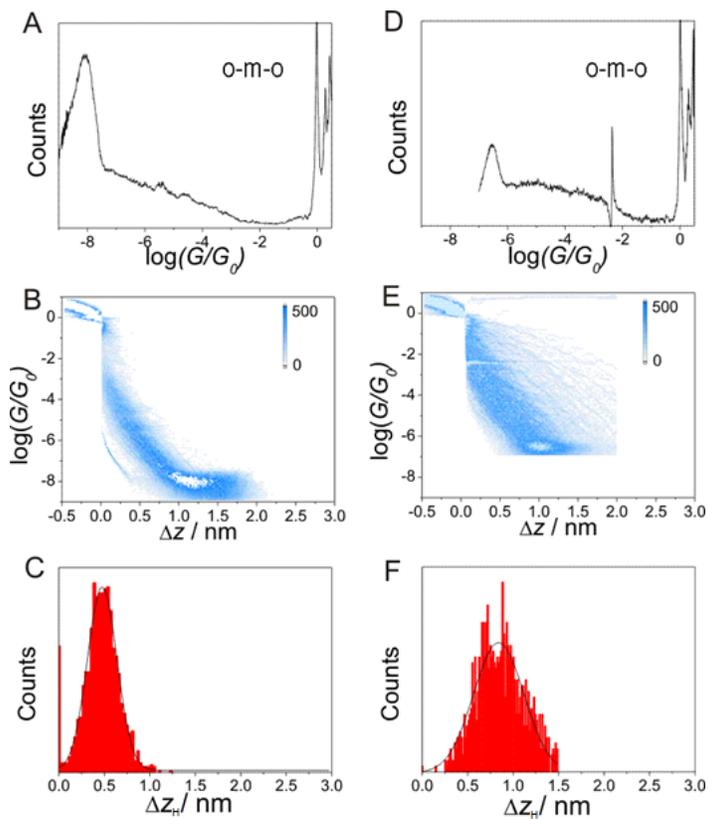

**Supplementary Figure 12.** 1D conductance histograms (A, D), 2D histograms (B, E), and distributions of the 'end-of-high-conductance-plateau displacements' $\Delta z_H$ (C, F) of o-m-o. Columns (A-C) were obtained using MCBJ and (D-F) using STM-BJ. For this molecule, conductance measurement provided pure tunneling traces. In this case, there are no discernible high-conductance plateaus and therefore, the peaks shown in C and F were assigned to pure tunneling. $G_H^{end}$ chosen to construct the $\Delta z_H$ distribution is $10^{-6.0}$ $G_0$ for MCBJ and $10^{-6.0}$ $G_0$ for STM-BJ.

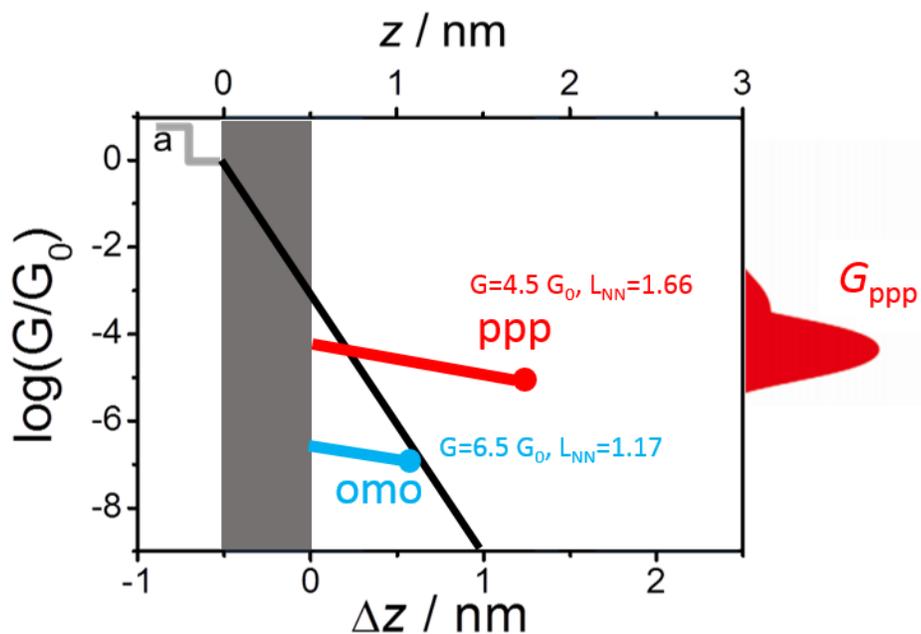

**Supplementary Figure 13.** A simple schematic to describe the parallel tunneling pathway. $\Delta z$ in the bottom axis describes the relative distance scale from the break junction measurement and $z$ in the upper axis describes the absolute distance scale between the two gold electrodes. The black line describes direct tunneling between gold electrodes in the absence of a molecular junction. The red and blue lines describe the expected conductance plateaus of p-p-p and o-m-o molecular junctions, respectively. Parameters are estimated from Table 1 and Supplementary Figure S12.

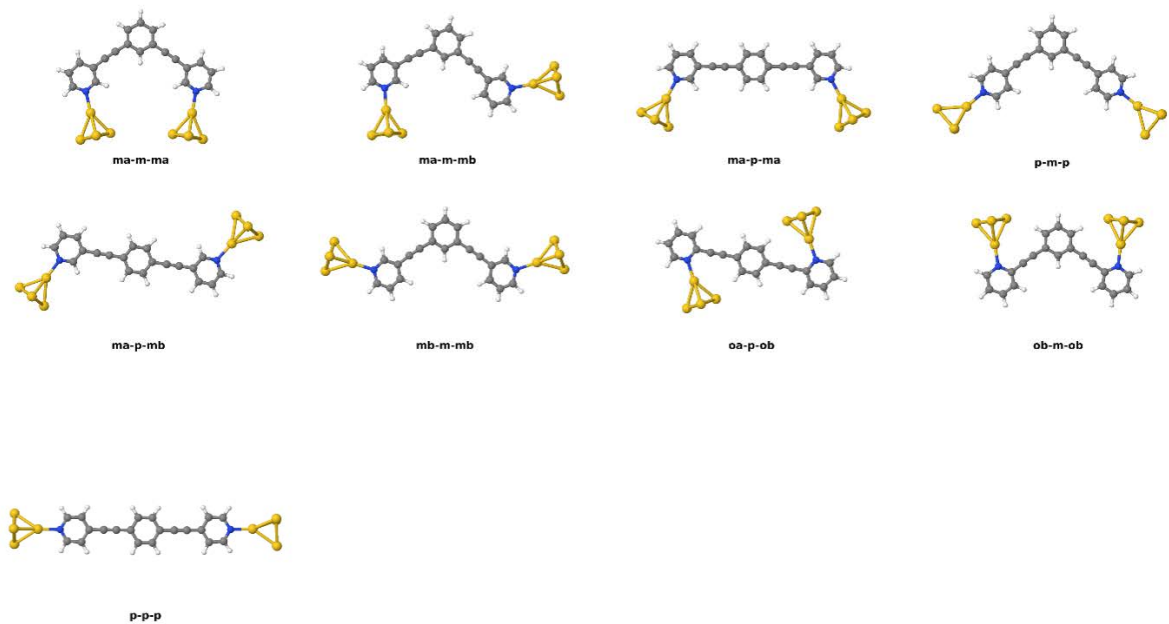

**Supplementary Figure 14.** Various idealized junctions with connected para, meta and ortho aromatic rings, illustrating the versatile planar conformations of the three-aromatic ring systems as possible components of molecular circuits.

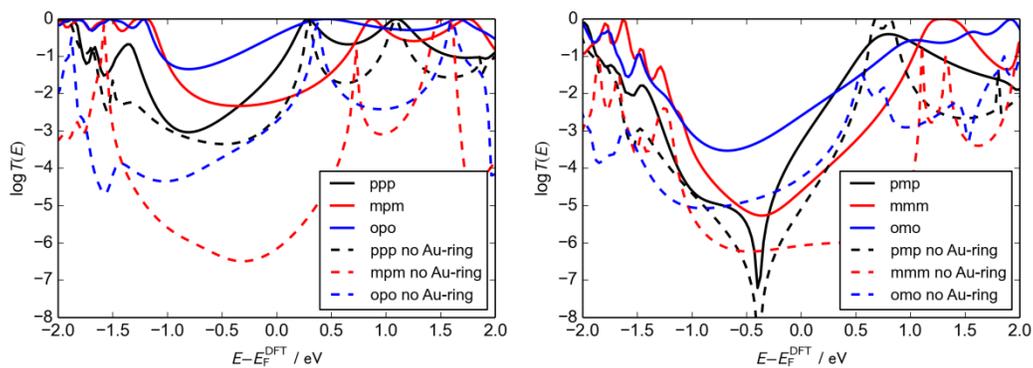

**Supplementary Figure 15.** Transmission coefficient functions for junction geometries shown at most left hand side in Supplementary Figure S16 and Supplementary Figure S17 for molecules in group 1 (left pane) and 2 (right pane), with ring-gold (continuous) and without ring-gold (dashed) coupling in the Hamiltonian matrix.

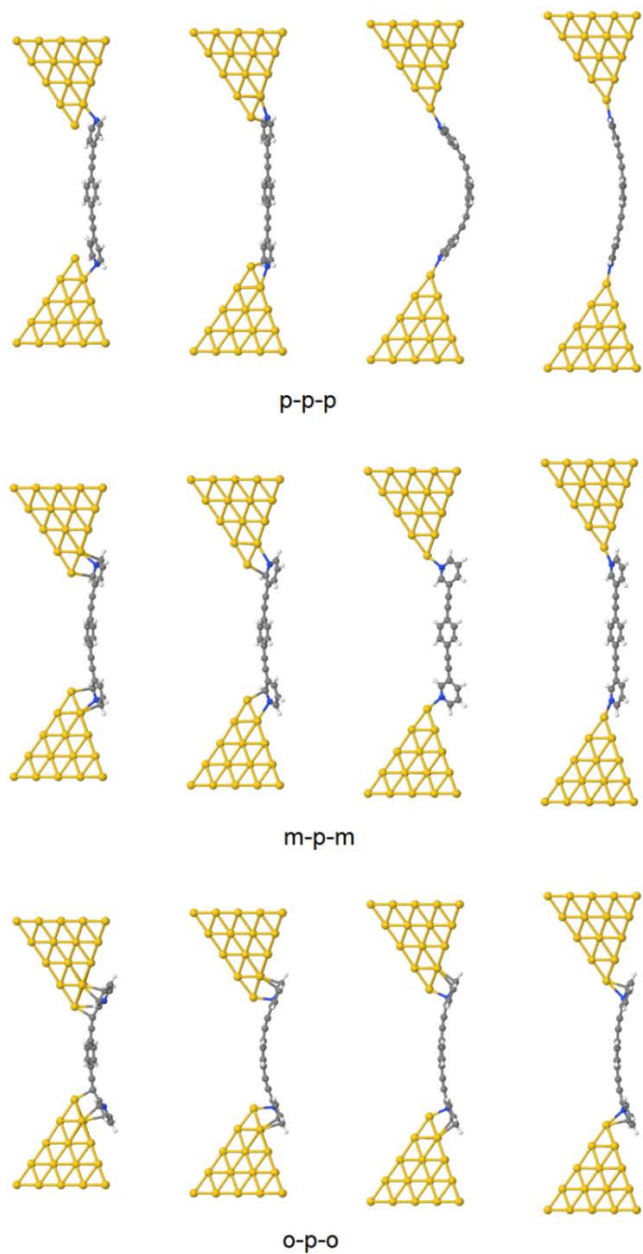

**Supplementary Figure 16.** Junction geometries for group 1 molecules in the BJ simulation. The electrode separation ($z_{the}$) increases from left to right. For each molecule there are four geometries that correspond to the four theoretical conductance points in the simulated trace in Supplementary Figure S18 in the main text.

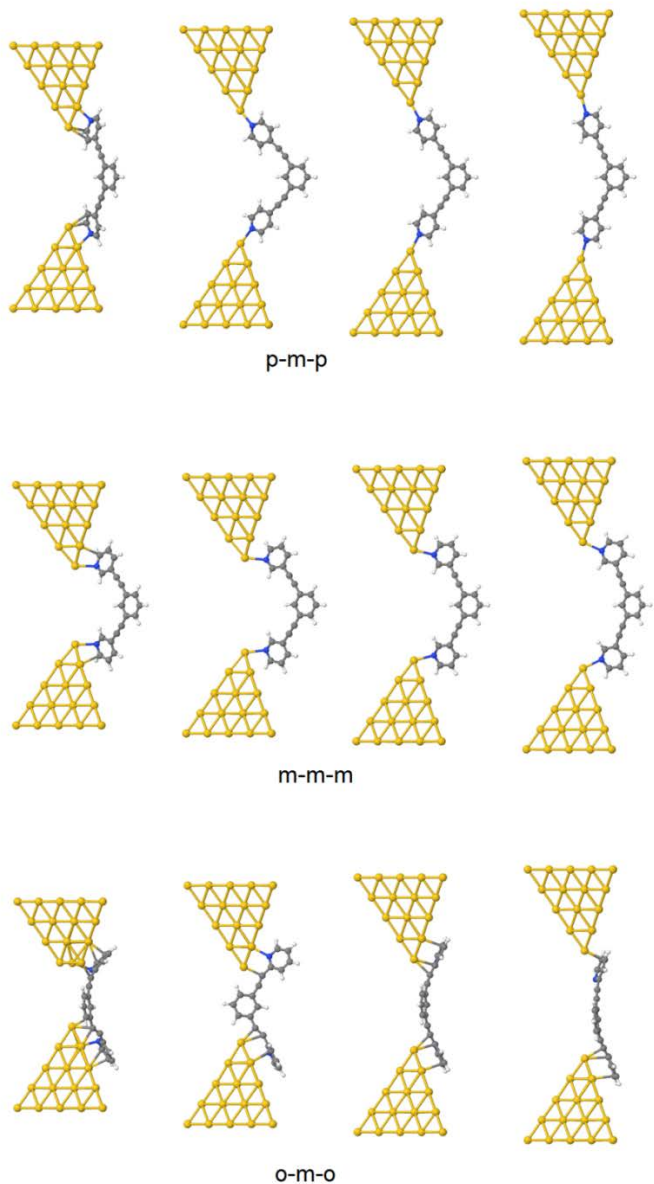

**Supplementary Figure 17.** Junction geometries for group 2 molecules in the BJ simulation. The electrode separation ($z_{the}$) increases from left to right. For each molecule there are four geometries that correspond to the four theoretical conductance points in the simulated trace in Supplementary Figure S18.

Supplementary Figure 18 shows the logarithm of conductance as a function of the theoretical electrode separation $z_{the}$, defined as $z_{the}=z_{Au-Au}-0.25$ nm, where $z_{Au-Au}$ is the tip separation in the relaxed structure, and 0.25 nm is the value of $z_{Au-Au}$ when the conductance through the two contacting pyramids (in the absence of a molecule) is $G_0$. Figure 18 also shows the experimental most-probable high conductance, plotted against $z_H*$ [6].

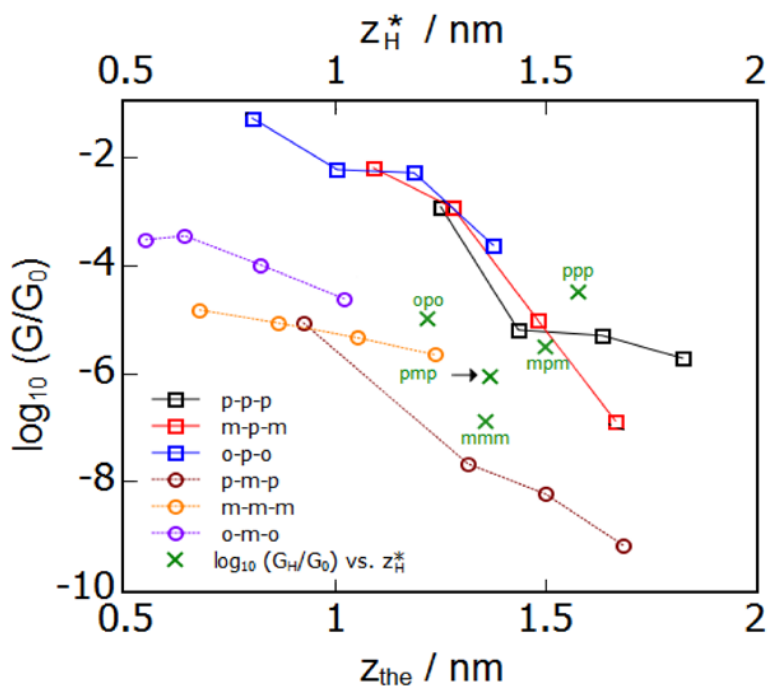

**Supplementary Figure 18.** Computed conductance vs electrode separation. Simulated trace curves for the first and second groups of molecules, marked with rectangle and circle markers, respectively. The trace curves show the logarithm of the conductance versus the theoretical electrode separation $z_{the}$ (bottom horizontal axis). Beyond the last points of the simulated trace curves, the simulated junction (i.e. the Au-N bond) is broken. Junction configurations for each stage during the stretching can be seen in Supplementary Figures 16-17. The green crosses are the experimental most-probable high conductance values plotted against the experimental values of $z_H*$. The simulated trace curves and the experimental conductance values demonstrate a distinct separation between groups 1 and 2.

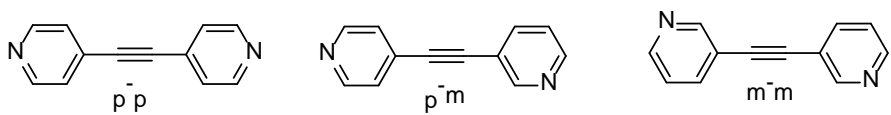

**Supplementary Figure 19**. Structures of the molecules studied theoretically in Supplementary Figure 20.

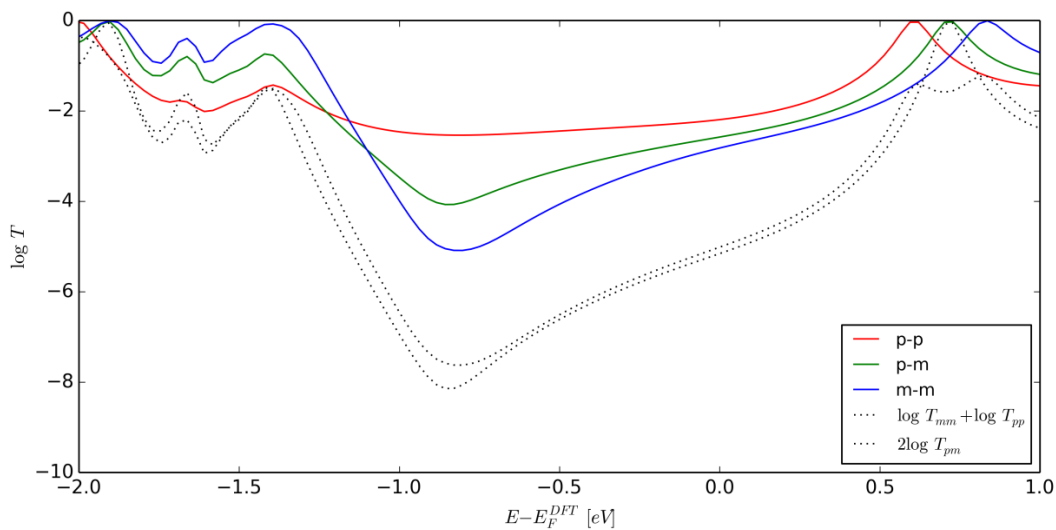

**Supplementary Figure 20.** Transmission coefficient function for the two pyridyl ring systems shown in Supplementary Figure 19.

Supplementary Note 1 Synthesis and Characterization of Molecules

Synthetic Procedures

All reactions were conducted under a blanket of argon which was dried by passage through a column of phosphorus pentoxide. All commercial chemicals were used without further purification. Anhydrous solvents were dried through an HPLC column on an Innovative Technology Inc. solvent purification system. Column chromatography was carried out using 40-60 μm mesh silica (Fluorochem). NMR spectra were recorded on: Bruker Avance-400, Varian VNMRS-600, VNMRS-700 and Varian Inova 500 spectrometers. Chemical shifts are reported in ppm relative to tetramethylsilane (0.00 ppm). Melting points were determined in open-ended capillaries using a Stuart Scientific SMP40 melting point apparatus at a ramping rate of 2 °C/min. Mass spectra were measured on a Waters Xevo OTofMS with an ASAP probe. Electron ionisation (EI) mass spectra were recorded on a Thermoquest Trace or a Thermo-Finnigan DSQ.

**General Procedure for Sonogashira Cross-Coupling Reactions:** To a solution of the iodoarene or bromoarene (1.0 eq), [Pd(PPh$_3$)$_4$] or [PdCl$_2$(PPh$_3$)$_2$] (3-5% mole) and CuI (3-5% mole) in THF/Et$_3$N (3:1 v/v) or THF/(*i*-Pr)$_2$NH (3:1 v/v) mixture (7 mL per 1 mmol of iodoarene or bromoarene), the alkyne was added and the mixture was stirred at 20 °C under argon for 6-36 h. The solvent was removed by vacuum evaporation and the product was purified by column chromatography using the eluent stated.

Scheme 1 Synthesis of P-P-P

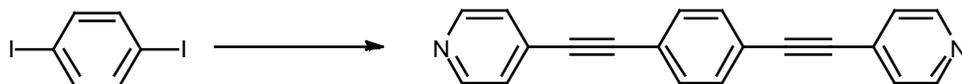

**Reagents and conditions**: 4-ethynylpyridine, [Pd(PPh$_3$)$_4$], CuI, THF/ Et$_3$N, 18 h, 50 °C, 64%.

**General Procedure**: 1,4-diiodobenzene (200 mg, 0.61 mmol), CuI (12 mg), [Pd(PPh$_3$)$_4$] (71 mg), 4-ethynylpyridine[1] (187 mg, 1.83 mmol), THF/Et$_3$N (12 ml), 18 h, 50 °C. Solvent was removed by vacuum evaporation. Purification by column chromatography using dichloromethane/ethyl acetate (3:2 v/v) as eluent gave **P-P-P** as a pale yellow solid (110 mg, 64% yield). m.p.: 194.8-196.2 °C. $^1$H NMR (400 MHz, CDCl$_3$) δ 8.62 (d, $J$ = 5.9 Hz, 4H), 7.54 (s, 4H), 7.36 (d, $J$ = 5.9 Hz, 4H). $^{13}$C NMR (101 MHz, CDCl$_3$) δ 150.12, 132.15, 131.20, 125.74, 123.10, 93.35, 89.08. MS (ASAP+) *m/z*: 281.1034 ([M+H]$^+$, 100%). The NMR spectroscopic data are in agreement with those in the literature.[2]

Scheme 2 Synthesis of O-P-O

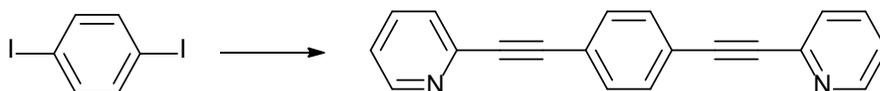

**Reagents and conditions**: 2-ethynylpyridine, [Pd(PPh$_3$)$_4$], CuI, THF/Et$_3$N, 18 h, 50 °C, 71%.

**O-P-O; General Procedure:** 1,4-diiodobenzene (200 mg, 0.61 mmol), CuI (12 mg), [Pd(PPh$_3$)$_4$] (71 mg), 2-ethynylpyridine (187 mg, 1.83 mmol), THF/Et$_3$N (12 ml), 18 h, 50 °C. Solvent was removed by vacuum evaporation. Purification by column chromatography using dichloromethane/ethyl acetate (3:2 v/v) as eluent gave **O-P-O** as a pale yellow solid (120 mg, 71% yield). m.p.: 190.8-192.3 °C. $^1$H NMR (400 MHz, CDCl$_3$) δ 8.63 (m, 2H), 7.69 (m, 2H), 7.58 (s, 4H), 7.53 (m, 2H), 7.25 (m, 2H). $^{13}$C NMR (101 MHz, CDCl$_3$) δ 150.39, 143.40, 136.44, 132.22, 127.51, 123.20, 123.04, 90.81, 88.83. Calcd for C$_{20}$H$_{12}$N$_2$: C, 85.69; H, 4.31; N, 9.99. Found: C, 85.53; H, 4.28; N, 9.97. HR-MS (ASAP+) *m/z* calcd for C$_{20}$H$_{12}$N$_2$ [M]$^+$ 280.1000, found *m/z* : [M]$^+$ 280.0982.

Scheme 3 Synthesis of M-P-M

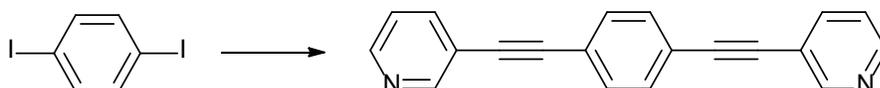

**Reagents and conditions**: 3-ethynylpyridine, [Pd(PPh$_3$)$_4$], CuI, THF/ Et$_3$N, 17 h, r.t., 76%.

**M-P-M; General Procedure:** 1,4-diiodobenzene (200 mg, 0.61 mmol), CuI (12 mg), [Pd(PPh$_3$)$_4$] (71 mg), 3-ethynylpyridine (187 mg, 1.83 mmol), THF/Et$_3$N (12 ml), 17 h, r.t.. Solvent was removed by vacuum evaporation. Purification by column chromatography using dichloromethane/ethyl acetate (1:9 v/v) as eluent gave **M-P-M** as a pale yellow solid (130 mg, 76% yield). m.p.: 190.1-193.0 °C. $^1$H NMR (400 MHz, CDCl$_3$) δ 8.77 (bs, 2H), 8.56 (m, 2H), 7.81 (m, 2H), 7.54 (s, 4H), 7.30 (m, 2H). $^{13}$C NMR (101 MHz, CDCl$_3$) δ 152.50, 149.05, 138.67, 131.92, 123.31, 123.10, 120.36, 92.34, 88.25. Calcd for C$_{20}$H$_{12}$N$_2$: C, 85.69; H, 4.31; N, 9.99. Found: C, 85.51; H, 4.25; N, 9.97. HR-MS (ASAP+) *m/z* calcd for C$_{20}$H$_{12}$N$_2$ [M]$^+$ 280.1000, found *m/z* : [M]$^+$ 280.1008.

Scheme 4 Synthesis of P-M-P

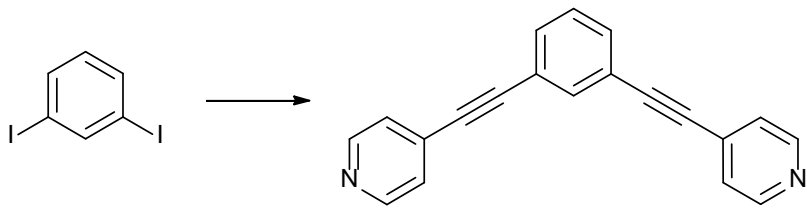

**Reagents and conditions**: 4-ethynylpyridine, [Pd(PPh$_3$)$_4$], CuI, THF/ Et$_3$N, 24 h, r.t., 88%.

**P-M-P; General Procedure:** 1,3-diiodobenzene (120 mg, 0.36 mmol), CuI (7 mg), [Pd(PPh$_3$)$_4$] (41 mg), 4-ethynylpyridine (112 mg, 1.02 mmol), THF/Et$_3$N (12 ml), 24 h, r.t.. Solvent was removed by vacuum evaporation. Purification by column chromatography using dichloromethane/ethyl acetate (1:1 v/v) as eluent gave **P-M-P** as a pale yellow solid (90 mg, 88% yield). m.p.: 132.8-134.2 °C. $^1$H NMR (400 MHz, CDCl$_3$) δ 8.58 (d, *J* = 4.8 Hz, 4H), 7.72 (t, *J* = 1.6 Hz, 1H), 7.52 (dd, *J* = 7.8, 1.6 Hz, 2H), 7.37 (s, 1H), 7.34 (d, *J* = 4.8 Hz, 4H). $^{13}$C NMR (101 MHz, CDCl$_3$) δ 150.04, 135.25, 132.57, 131.19, 129.01, 125.74, 122.90, 92.83, 87.69. Calcd for C$_{20}$H$_{12}$N$_2$: C, 85.69; H, 4.31; N, 9.99. Found: C, 85.48; H, 4.27; N, 9.93. HR-MS (ASAP+) *m/z* calcd for C$_{20}$H$_{12}$N$_2$ [M]$^+$ 280.1000, found *m/z* : [M]$^+$ 280.0985.

Scheme 5 Synthesis of M-M-M

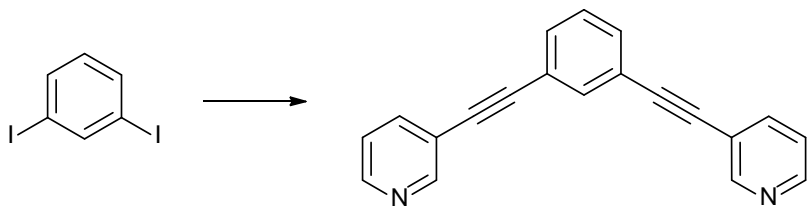

**Reagents and conditions**: 3-ethynylpyridine, [Pd(PPh$_3$)$_4$], CuI, THF/ Et$_3$N, 17 h, r.t., 58%.

**M-M-M; General Procedure:** 1,3-diiodobenzene (200 mg, 0.61 mmol), CuI (12 mg), [Pd(PPh$_3$)$_4$] (71 mg), 3-ethynylpyridine (85.8 mg, 1.32 mmol), THF/Et$_3$N (12 ml), 17 h, r.t.. Solvent was removed by vacuum evaporation. Purification by column chromatography using dichloromethane/ethyl acetate (1:9 v/v) as eluent gave **M-M-M** as a pale yellow solid (100 mg, 58% yield). m.p.: 84.0-85.7 °C. $^1$H NMR (400 MHz, CDCl$_3$) δ 8.74 (d, *J* = 1.5 Hz, 2H), 8.52 (dd, *J* = 4.9, 1.7 Hz, 2H), 7.76 (m, 2H),

7.70 (m, 1H), 7.49 (m, 2H), 7.33 (m, 1H), 7.25 (m, 2H). $^{13}$C NMR (101 MHz, CDCl$_3$) δ 152.48, 149.00, 134.92, 132.03, 128.90, 123.29, 123.23, 120.31, 91.78, 86.96. Calcd for C$_{20}$H$_{12}$N$_2$: C, 85.69; H, 4.31; N, 9.99. Found: C, 85.57; H, 4.29; N, 9.92. HR-MS (ASAP+) *m/z* calcd for C$_{20}$H$_{12}$N$_2$ [M]$^+$ 280.1000, found *m/z* : [M]$^+$ 280.1011.

Scheme 6 Synthesis of O-M-O

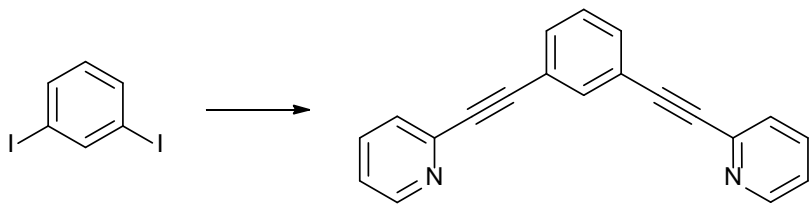

**Reagents and conditions**: 2-ethynylpyridine, [Pd(PPh$_3$)$_4$], CuI, THF/ Et$_3$N, 20 h, 50 °C, 52%.

**O-M-O; General Procedure:** 1,3-diiodobenzene (250 mg, 0.76 mmol), CuI (14 mg), [Pd(PPh$_3$)$_4$] (88 mg), 2-ethynylpyridine (234 mg, 2.28 mmol), THF/Et$_3$N (12 ml), 20 h, 50 °C. Solvent was removed by vacuum evaporation. Purification by column chromatography using dichloromethane/ethyl acetate (1:9 v/v) as eluent gave **O-M-O** (110 mg, 52% yield). m.p.: 72.3- 73.8 °C. $^1$H NMR (400 MHz, CDCl$_3$) δ 8.66 (m, 2H), 7.85 (m, 1H), 7.72 (td, *J* = 7.7, 1.8 Hz, 2H), 7.62 (dd, *J* = 7.8, 1.6 Hz, 2H), 7.56 (dt, J = 7.8, 1.1 Hz, 2H), 7.39 (t, J = 8.1 Hz, 1H), 7.28 (m, 2H). $^{13}$C NMR (101 MHz, CDCl$_3$) δ 150.29, 143.29, 136.41, 135.42, 132.58, 128.84, 127.50, 123.19, 122.92, 89.51, 88.20. Calcd for C$_{20}$H$_{12}$N$_2$: C, 85.69; H, 4.31; N, 9.99. Found: C, 85.57; H, 4.29; N, 9.92. HR-MS (ASAP+) *m/z* calcd for C$_{20}$H$_{13}$N$_2$ [M+H]$^+$ 281.1079, found *m/z* : [M+H]$^+$ 281.1059.

**Molecular properties: NMR Spectra**

## Supplementary Note 2 Molecular electronic properties

Supplementary Figures S7-S12 show the molecular electronic properties for the molecules from MCBJ and STM-BJ techniques in detail

**Supplementary Note 3  The limitation of break junction technique to determine single-molecule conductance of o-m-o molecule**

According to quantum circuit rule, the conductance of the o-m-o molecule is expected to be $10^{-6.5}$ $G_0$, which is within our sensitivity limit (~$10^{-8}$ $G_0$) of MCBJ set up. However, as shown in Supplementary Figure12, we are not able to determine the molecular conductance feature of o-m-o molecular junction. To confirm this experimental finding, we repeated the conductance measurement more than ten times and always get similar results. Our further analysis reveal that the absence of conductance feature in our break junction measurement is actually because of the combination effect of the relative short N---N length and low conductance, which leads to the phenomenon that the direct tunneling conductance between the two gold electrodes covers the single-molecule conductance.

According to the quantum circuit rule, the conductance of the o-m-o molecule is expected to be $10^{-6.5}$ $G_0$, which is within the sensitivity limit (~$10^{-8}$ $G_0$) of our MCBJ set up. However, as shown in Supplementary Figure12, we are not able to determine the molecular conductance feature of o-m-o molecular junction. To confirm this experimental finding, we repeated the conductance measurement more than ten times and always get similar results. Our further analysis reveals that the absence of a conductance feature in our break junction measurement is actually because of the combination effect of the relatively short N---N length and low conductance, which leads to the phenomenon that the direct tunneling conductance between the two gold electrodes obscures the single-molecule conductance.

Supplementary Figure13 demonstrates a simple schematic of typical $G$-$\Delta z$ traces in break junction. After breaking of the gold-gold contact (shown as step "a" at $G_0$), the tunneling conductance decreases exponentially as the black line and the slope was estimated from the experimental data. The gray area indicates the tunneling conductance changes within the snap-back area, which cannot be monitored from the break junction measurement. When we have a molecule junction between the two gold leads, the tunneling conductance through the molecular junction can be higher than the direct tunneling through solvent, for instance, for a p-p-p molecule we are able to determine a clear molecular conductance plateau at around $10^{-4.5}$ $G_0$ (red curve, the parameter is estimated from data shown in Table 1), which contributed to a clear conductance peak in the conductance histogram after statistical analysis. However, for o-m-o, the direct tunneling conductance with the associated N---N distance of the o-m-o molecule (which is calculated to be 1.17 nm shown in Table 1) is ~$10^{-6}$ $G_0$, which is slightly higher than the expected conductance of the o-m-o single-molecule junction (which is expected to be ~$10^{-6.5}$ $G_0$). In this case, the expected molecular conductance plateau (blue line) will be completely covered by direct tunneling conductance, to the extent that the conductance peak cannot appear in the conductance histogram.

A similar phenomenon was also observed and discussed in a previous study[3], and the case of o-m-o can be considered as an exception in the break junction measurement.

**Supplementary Note 4 Differences between MCBJ and STM-BJ results**

In this section, we comment on differences which can arise between MCBJ and STM-BJ measurements. After establishing the MCBJ and STM-BJ set-ups in Bern, we have carried three molecular system studies of tolane, oligophenylene and oligoyne with different anchoring groups to compare the results from MCBJ and STM-BJ technique[4-6]. In most cases, the two techniques showed nice agreement, but sometimes give slightly different results. Differences in the measurements using MCBJ and STM-BJ techniques can arise from:

1. The sensitivity of current measurement: for our MCBJ set-up, we construct a highly sensitive log-amplifier that we can measure current down to 10 fA range. Benefiting from this advantage, as shown in the 1D and 2D conductance histograms, in MCBJ we can observe low-conductance features in most cases, while in our STM-BJ measurements this feature is covered by noise of the linear amplifier around 1 pA.

2. Mechanical stability: MCBJ technique is expected to provide higher mechanical stability than STM-BJ technique. The mechanical stability may partially influence the dynamics of molecules between the two gold-electrodes, which may lead to some minor conductance difference, especially for molecules with relative weak binding.

3. Stretching rate: due to the different current amplifier design on our STM-BJ and MCBJ and different mechanical stability, we applied different stretching rates for our STM-BJ at ~60 nm/s and MCBJ at ~5 nm/s. The relative slow rate of MCBJ experiment is to take advantage of good mechanical stability and to follow the relative slow response of the log-amplifier in fA current region. The stretching rate difference may also lead to some slight change in the junction behaviors.

4. Experimental variation: Practically we have found the conductance variation in break junction experiments (STM-BJ/MCBJ) can be up to 0.1-0.2 in log scale and stretching distance variation can be up to 0.2 nm from experiment to experiment even in well-defined conditions.

Supplementary Note 5 Idealized junction geometries

These are shown in Supplementary Figure14

## Supplementary Note 6  Break-Junction simulations

To further analyze BJ events we performed DFT based BJ simulations by inserting the molecule between two opposing 35 atom gold (111) directed pyramids with various tip separations shorter than the molecular lengths and performed a geometry optimization of the structure. (The tip separation is defined as the center-to-center distance between the apex atoms of the two opposing pyramids.) The relaxed junction geometries can be seen in Supplementary Figure16 and Figure17. To produce conductance-trace curves, the transmission coefficient $T(E)$ describing electrons of energy $E$ passing from one electrode to the other via the molecule was calculated for each tip separation and the conductance $G/G_0 = T(E_F)$ was obtained by evaluating $T(E)$ at the Fermi energy $E_F$. A few transmission coefficient functions can be seen in Supplementary Figure15

Removing the ring-gold coupling generally reduces the conductance as the dominant resonances become narrower. The primary signature of destructive interference in the cases of meta terminal and central rings is the low value of the conductance. The anti-resonance in the pi channel is not always evident, because it may be masked by parallel sigma-like channels[7,8].

Figure S18 shows that for any given value of $z_{the}$ or $z_H*$, both the theoretical and experimental conductances of the molecules in group 1 are distinctly higher than those of the molecules in group 2. Supplementary Figure S18 also shows that for many values of $z_{the}$ (in the electrode separation range 1.0 nm<$z_{the}$<1.4 nm) the theoretical conductances of molecules within group 1 are rather similar and therefore we conclude that QI in the central ring is more important than QI in the anchors. For molecules in group 2, the theoretical conductance for m-m-m is larger than for p-m-p, which is in the reverse order of the experimental most probable high conductance values. The latter artifact is attributed to unrealistically-large metal-ring coupling in the m-m-m simulated optimized junction geometries that will rarely occur in room temperature environments.

## Supplementary Note 7 Phase averaging in an ensemble of measurements

Since the experimental conductance values are of statistical origin, a product rule for ensemble averaged conductances can arise due to inter-ring phase averaging. To illustrate this point, consider two quantum scatterers (labelled 1 and 2) in series, whose transmission and reflection coefficients are $T_1$, $T_2$ and $R_1$, $R_2$ respectively. It can be shown that the total transmission coefficient for the scatterers in series is $T = \frac{T_1 T_2}{1 - 2\sqrt{R_1 R_2} \cos \varphi + R_1 R_2}$, where φ is a quantum phase arising from QI between the scatterers[9,43]. We note that the experimental quoted conductance is identified with the most probable value of $\log_{10} \frac{G}{G_0}$ and if this possesses a Gaussian distribution, then it equates to the ensemble average of $\log_{10} \frac{G}{G_0}$, which we denote by $\overline{\log_{10} T}$. From the above expression for T,

$$\overline{\log_{10} T} = \overline{\log_{10} T_1} + \overline{\log_{10} T_2} - \overline{\log_{10}(1 - 2\sqrt{R_1 R_2} \cos \varphi + R_1 R_2)}$$

Although each individual molecule is phase coherent, if the ensemble average involves an average over the phase, uniformly distributed between 0 and $2\pi$, then the third term on the right hand side averages to zero, because of the mathematical identity $\int_0^{2\pi} d\varphi \, \overline{\log_{10}(1 - 2\sqrt{R_1 R_2} \cos \varphi + R_1 R_2)} = 0$. Hence in the ensemble average, all information about the inter-scatterer quantum phase is lost and $\overline{\log_{10} T} = \overline{\log_{10} T_1} + \overline{\log_{10} T_2}$, or equivalently $\frac{G_{Total}}{G_0} = \frac{G_1}{G_0} \frac{G_2}{G_0}$.

For the three-ring OPEs of interest, if inter-ring QI is similarly absent in the ensemble average, then one would expect

$\frac{G_{Tot}}{G_0} = \frac{G_t}{G_0} \frac{G_c}{G_0} \frac{G_t}{G_0}$, where $G_t$ is the conductance of the terminal pyridyl ring and $G_c$ is the conductance of the central phenyl ring. In the absence of inter-ring QI, the above expression implies that the 'quantum circuit rule' is satisfied.

## Supplementary Note 8 A circuit rule for two rings.

For further demonstrate the generality of the product rule we performed DFT based transport calculations (with the exact same methodology as gave the result in Figure 4) for two pyridyl ring systems with para and meta connections. The structures of three molecules studied are shown in Supplementary Figure S19. The theoretical derivation above implies that for two rings

$$G_{pp} G_{mm} = G_{pm}^2,$$

Where the $G_{pp}$ and $G_{mm}$ are the conductances of para-para and meta-meta pyridyl rings, and $G_{mp}$ is the conductance of the molecule for meta and para pyridyl rings. Supplementary Figure S20 demonstrates that the product rule is satisfied for a wide range of energies (the dotted curves match).

**Supplementary note 9 Decoherence in a fully-coherent theory**

It may seem surprising that phase averaging can lead to decoherence, even though the theory describing transport is fully phase coherent. Decoherence is perfectly reconciled with the calculations in Figure 4 and Figure 5, because although individual electrons remain phase coherent over the whole molecule, a restricted form of decoherence can arise in the ensemble averages of transport properties. This is important, because all break-junction experiments involve averages over large ensembles in a single measurement and the typical conductance associated with a compound is the most occurred value of log(G/G_0) for several measurements. For a complete discussion we show a possibility that our conductance product rule could arise from not only the intrinsic electronic properties of the individual molecules, but from the statistical origin of the experimental conductance values. We have modified the manuscript such that this argument now is clear.

Perhaps our discussion can be made clearer through an analogy. In the NMR literature, the phrase 'decoherence' is often used to describe two distinct phenomena, which are assigned "T1" and "T2" coherence times. The latter does not require inelastic scattering and arises when spins within an ensemble precess in the plane perpendicular to the applied magnetic field with slightly different frequencies. Consequently, even in the absence of inelastic process and even though individual spins remain coherent, the ensemble of spins eventually loses coherence. In contrast T1 processes involve spin flips. These are inelastic processes and provide a second source of decoherence. In our work, all transport is elastic and individual electrons remain perfectly coherent. Nevertheless for an ensemble of measurements on such electrons, decoherence in average values can arise from random elastic scattering, just as decoherence in NMR experiments arises from 'T2' processes.

**Supplementary references**